\documentclass[linenumbers]{aastex631}
\usepackage{multirow} 
\usepackage{makecell}
\usepackage{diagbox}
\usepackage{booktabs}
\usepackage{color}
\usepackage{tabularx}
\usepackage{rotating}
\newcommand{\dmunit}{ {$\rm pc\hspace{0.24em} cm^{-3}$} }
\newcommand{\hunit}{ {$\rm km \hspace{0.24em} s^{-1} \hspace{0.24em}Mpc^{-1}$} }
\newcommand{\bcwidth}{$\Delta t_{bc}$}
\newcommand{\widthfitb}{$\Delta t_{fitb}$}
\newcommand{\flux}{$S_{\nu}$}
\newcommand{\fluence}{$F_{\nu}$}
\newcommand{\scattime}{$\Delta t_{sc}$}
\newcommand{\spidx}{$\gamma$}
\newcommand{\sprun}{$r$}
\newcommand{\highfreq}{$\nu_{max}$}
\newcommand{\lowfreq}{$\nu_{min}$}
\newcommand{\peakfreq}{$\nu_{p}$}
\newcommand{\rw}{$\Delta t_{rw}$}
\newcommand{\bt}{$T_{B}$}

\begin{document}
\nolinenumbers
\title{Unsupervised Machine Learning for Classifying CHIME Fast Radio Bursts and Investigating Empirical Relations}

\author{Da-Chun Qiang}\thanks{e--mail: dcqiang@hnas.ac.cn}
\affiliation{Institute for Gravitational Wave Astronomy, Henan Academy of Sciences, Zhengzhou 450046, Henan, China}

\author{Jie Zheng}
\affiliation{Institute for Gravitational Wave Astronomy, Henan Academy of Sciences, Zhengzhou 450046, Henan, China}

\author{Zhi-Qiang You}
\affiliation{Institute for Gravitational Wave Astronomy, Henan Academy of Sciences, Zhengzhou 450046, Henan, China}

\author{Sheng Yang}\thanks{Corresponding author; e--mail: sheng.yang@hnas.ac.cn}
\affiliation{Institute for Gravitational Wave Astronomy, Henan Academy of Sciences, Zhengzhou 450046, Henan, China}
\affiliation{INAF Osservatorio Astronomico di Padova, Vicolo dell’Osservatorio 5, I-35122 Padova, Italy}

\begin{abstract}
Fast Radio Bursts (FRBs) are highly energetic millisecond-duration astrophysical phenomena typically categorized as repeaters or non-repeaters. However, observational limitations may result in misclassifications, potentially leading to a higher proportion of repeaters than currently identified. In this study, we leverage unsupervised machine learning techniques to classify FRBs using data from the CHIME/FRB catalogs, including both the first catalog and a recent repeater catalog. By employing Uniform Manifold Approximation and Projection (UMAP) for dimensionality reduction and clustering algorithms (k-means and Hierarchical Density-Based Spatial Clustering of Applications with Noise [HDBSCAN]), we successfully segregate repeaters and non-repeaters into distinct clusters, identifying over 100 potential repeater candidates. Our analysis reveals several empirical relations within the clusters, including the ${\rm log \,}\Delta t_{sc} - {\rm log \,}\Delta t_{rw}$, ${\rm log \,}\Delta t_{sc} - {\rm log \,}T_B$, and $r - \gamma$ correlations, where ${\Delta t_{sc}, \Delta t_{rw}, T_B, r, \gamma}$ represent scattering time, rest-frame width, brightness temperature, spectral running, and spectral index, respectively. The Chow test results reveal that while some repeaters and non-repeaters share similar empirical relationships, the overall distinctions between the two groups remain significant, reinforcing the classification of FRBs into repeaters and non-repeaters. These findings provide new insights into the physical properties and emission mechanisms of FRBs. This study demonstrates the effectiveness of unsupervised learning in classifying FRBs and identifying potential repeaters, paving the way for more precise investigations into their origins and applications in cosmology. Future improvements in observational data and machine learning methodologies are expected to further enhance our understanding of FRBs.

\end{abstract}

\section{Introduction}\label{sec:intro}

Fast Radio Bursts (FRBs) are highly energetic astronomical phenomena characterized by millisecond-duration emissions. The first FRB signal was discovered in 2007 by \cite{Lorimer:2007qn}, and their existence was firmly established in 2013 when \cite{Thornton:2013iua} published observations of four similar events detected by the Australian Parkes Radio Telescope. Since then, FRBs have drawn significant attention in both astronomy and cosmology \citep[e.g.,][]{Lorimer:2018rwi, Keane:2018jqo, Petroff:2021wug, Xiao:2021omr, Xiao:2021viy, Xiao:2022bbj, Zhang:2013lta, Zhang:2018gjb, Zhang:2020ass, Zhang:2021kdu, Wang:2020aut, Wang:2020rkl, Wang:2019sio, Wang:2022ami, Gao:2014iva, Qiang:2020vta, Qiang:2021bwb}. However, the origin of FRBs remains unknown.

To explore their origin, numerous radio telescopes have dedicated time and efforts on FRBs, such as, the Deep Synoptic Array \citep[DSA;][]{DSA-10, DSA-2000}, Arecibo \citep{Spitler:2014fla}, Parkes \citep[e.g.,][]{Lorimer:2007qn, Burke-Spolaor:2014rqa, Petroff:2015bua, Ravi:2014gxa}, the Canadian Hydrogen Intensity Mapping Experiment \citep[CHIME;][]{CHIMEFRB-IM}, the Five-hundred-meter Aperture Spherical Radio Telescope \citep[FAST;][]{FAST-IM}, and the Australian Square Kilometer Array Pathfinder \citep[ASKAP;][]{ASKAP-IM}. So far, nearly a thousand FRBs have been observed \citep{Petroff:2016tcr, CHIMEFRBcat1, Jankowski:2023sot, Xu:2023did}, with nearly 100 of them have known redshifts \citep[e.g.][]{DSA-110_cat1, DSA-110_cat2, Bhardwaj2024ApJ, Gordon2023ApJ}. To date, the largest FRB sample is the first CHIME/FRB catalog \citep{CHIMEFRBcat1}.

Based on the observational characteristics, FRBs are broadly classified into two categories, i.e.,  repeaters and non-repeaters. As the names suggest, repeaters are sources that have exhibited multiple bursts, whereas non-repeaters have only been observed to burst once. Since the discovery of the first repeater, FRB\,20121102 \citep{Spitler:2016dmz}, more than 50 FRBs have been identified as repeaters to date \citep{ CHIMEFRBcat1, CHIMEFRB2023, Kumar2019ApJ, Kirsten:2021llv, Fonseca:2020cdd, Niu:2021bnl, Xu2022Natur}, while most of the remaining ones are classified as non-repeaters. However, \cite{Ravi:2019iop} suggests that the volumetric rate of non-repeating FRBs might exceed that of cataclysmic events and the formation rate of compact objects, implying that the majority of FRBs should be repeaters. Some studies also propose that more than half of the FRBs in the first CHIME/FRB catalog could be repeaters \citep{Yamasaki:2023dlb, McGregor:2023mzr}. Furthermore, several FRBs initially identified as non-repeaters were later observed to repeat \citep{CHIMEFRB2023}. These suggest that many of the FRBs currently classified as non-repeaters might actually be potential repeaters, with only a single burst detected due to various observational factors. As a result, some studies have attempted to identify potential repeaters among apparent non-repeating FRBs, and one of the methods being employed is machine learning.

Machine learning is an artificial intelligence technique that allows computers to learn from data and make predictions or decisions without being explicitly programmed \citep{ml1967, ml1986}. Machine learning is generally categorized into three types, i.e., supervised learning, unsupervised learning, and semi-supervised learning \citep[e.g.][]{ml1, ml2, ml3, ml4}. Using algorithms to uncover patterns in the data can be applied to tasks such as classification, regression, clustering, and optimization. Currently, machine learning has already been widely used in the detection and analysis of FRBs \citep[e.g.][]{Wagstaff2016PASP, Zhang2018ApJ, Wu2019ApJ, Yang2021MNRAS, Adamek2020ApJS, Agarwal2020MNRAS, Bhatporia2023}. 
For instance within the first CHIME/FRB catalog dataset, \cite{Chen:2021jpq} and \cite{Zhu-Ge:2022nkz} identify 188 and 117 repeater candidates from 474 non-repeating FRBs via unsupervised learning, correspondingly.
\cite{Yang:2023dcf} instead discovered 145 repeater using FRB morphology as features, while \cite{Luo:2022smj} identified dozens of repeater candidates with various supervised learning methods. 
Furthermore, \cite{Luo:2022smj} found that the most prominent factors to distinguish between non-repeating and repeating FRBs are brightness temperature and rest-frame frequency bandwidth, whereas \cite{Sun:2024huw} found spectral running may play a role instead. Furthermore, some studies have used machine learning techniques to classify thousands of bursts from highly active repeaters, such as FRB\,20121102 \citep{Raquel:2023tgi} and FRB\,20201124A \citep{Chen:2023pub}, in an effort to analyze their potential radiation mechanisms.

In addition to classifying FRBs based on their repeatability, some studies have also investigated the clustering of FRBs using characteristics beyond repeatability, as well as the empirical relationships among these features. For instance, similar to the well-known classification of GRB, where short GRBs exhibit short prompt emissions and are mostly associated with the old population, while long GRBs with long prompt emissions and are mostly associated with the young population \citep{grb1, grb2}, \cite{Guo:2022wpf} proposed classifying FRBs based on their association with either old or young stellar populations. They discovered several tight empirical relations for non-repeaters in the first CHIME/FRB catalog, such as ${\rm log\,}E - {\rm log\,}L_{\nu}$, ${\rm log\,}E - {\rm log\, DM_E}$ and ${\rm log\, DM_E} - {\rm log\,}L_{\nu}$, where ${\rm log\,}$ means the logarithm to base 10, and $E$, ${\rm DM_E}$, $L_{\nu}$ are isotropic energy, spectral luminosity, and extragalactic DM, respectively. Similar empirical relations were found for non-repeaters associated with old populations and all non-repeaters, though with notably different slopes and intercepts, such as ${\rm log\,}F_{\nu} - {\rm log\,}S_{\nu}$, ${\rm log\, DM_E} - {\rm log\,}S_{\nu}$, and ${\rm log\,}F_{\nu} - {\rm log\, DM_E}$, where $F_{\nu}$ is specific fluence, $S_{\nu}$ is flux. Many empirical relations still hold for localized FRBs \citep{Li2024arXiv240812983}. Based on these empirical relations, FRBs could potentially serve as standard candles, allowing cosmological models to be constrained without relying on DM measurements \citep{Guo:2023hgb}. \cite{Li:2021yds} classified FRBs into short ($W < 100\,$ms) and long ($W > 100\,$ms) bursts, where $W$ represents the pulse width. A strong power-law correlation between fluence and peak flux density was identified for these categories. \cite{Xiao:2021viy} classified repeating FRBs into classical ($T_B \geq 10^{33} \, $K) and atypical ($T_B < 10^{33} \,$K) bursts, where $T_B$ refers to the brightness temperature. A tight power-law correlation between pulse width and fluence was also observed for classical bursts. 

CHIME has recently released a new repeater catalog \citep{CHIMEFRB2023}, significantly increasing the data available on repeaters. This expanded dataset is expected to improve the accuracy of identifying repeater candidates through machine learning. In this paper, we will apply unsupervised learning methods to classify the FRBs from the combined data of these two catalogs into different clusters, find potential repeaters among the non-repeaters, and analyze possible empirical relations across these clusters. This paper is structured as follows. In Section \ref{sec:data}, we present the selected CHIME data and the features used in our analysis. Section \ref{sec:met} introduces two types of unsupervised machine learning methods, i.e., dimensionality reduction and clustering, as well as evaluation metrics. In Section \ref{sec:res}, we present the results of dimensionality reduction and clustering, then we analyze the potential empirical relations. Finally, Section \ref{sec:con} provides our conclusions.

\section{DATA SET} \label{sec:data}

\subsection{Sample Construction} \label{subsec:sel}
In this paper, we use the largest available FRB data from CHIME to date, including the first CHIME/FRB Catalog \citep[][hereafter Cat1]{CHIMEFRBcat1} and the CHIME/FRB Collaboration (2023) Catalog \citep[][hereafter Cat2023]{CHIMEFRB2023}.
The Cat1 includes 536 events (474 non-repeaters and 62 repeat bursts from 18 repeaters). There are 600 sub-bursts in total because of multiple peaks appearing in the light curve of FRB.
The Cat2023 contains 127 events from 39 repeaters\,\footnote{In fact, \cite{CHIMEFRB2023} marked 14 FRB sources as repeater candidates due to lower significance, indicating that the burst-to-burst DM and sky position differences are larger compared to other confirmed repeaters, and their repetition rates are relatively low. In this study, we consider these 14 FRB sources as repeaters, as they warrant further follow-up observations for potential confirmation.}, or 151 sub-bursts for all events. Since there are 6 FRBs in Cat1 with flux and fluence values of zero, we excluded those FRBs. Additionally, 6 non-repeaters in Cat1, identified as repeaters, are duplicates in Cat2023, resulting in 739 FRB bursts used in this paper.

\subsection{Feature Selection} \label{subsec:fea}

\begin{figure}
    \centering
    \includegraphics[width=1.0\linewidth]{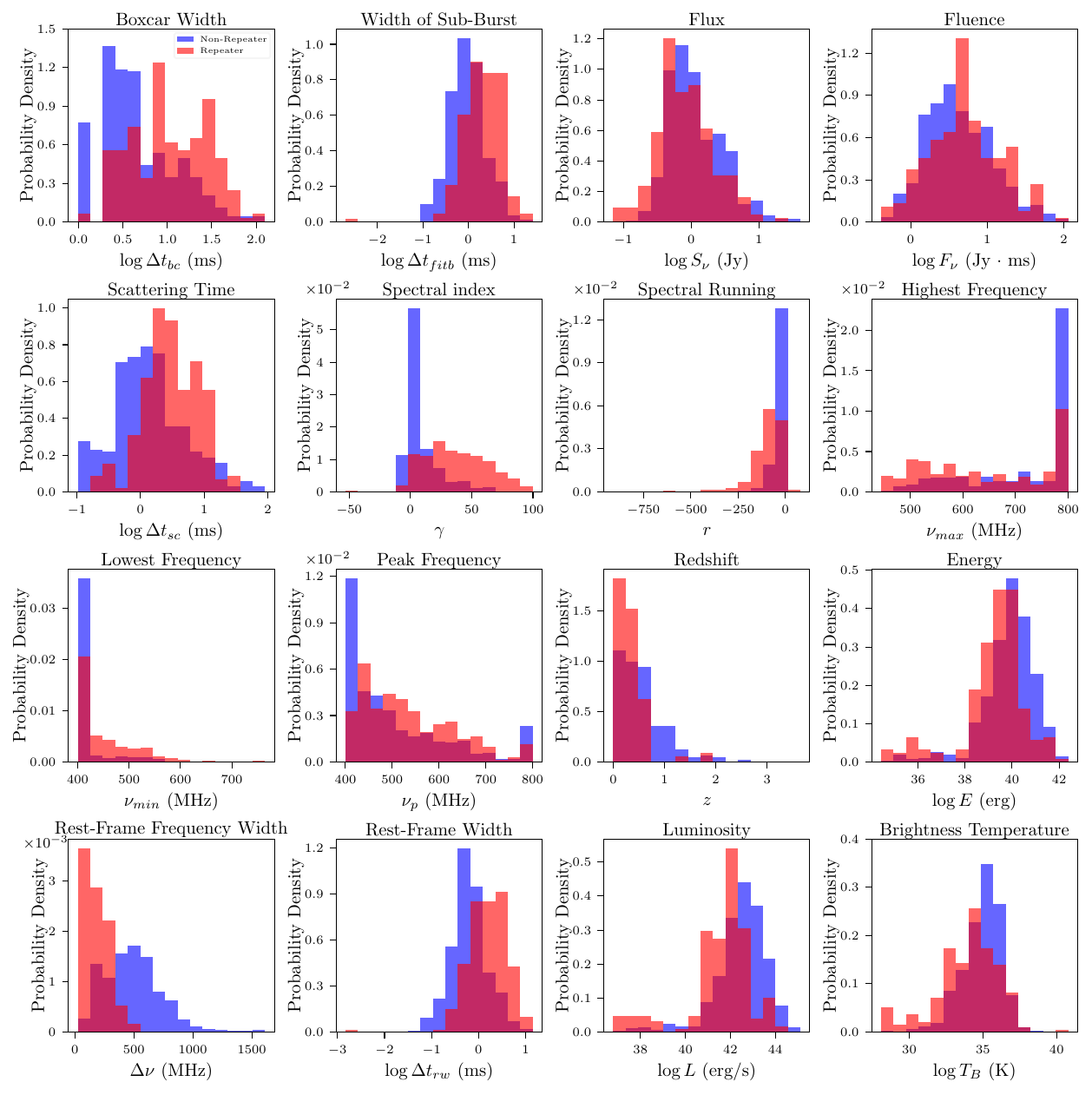}
    \caption{\label{fig:fd} The distributions of observed and derived parameters for non-repeaters and repeaters are shown separately, forming the input data used for unsupervised learning. For more details, refer to Section \ref{sec:data}.}

\end{figure}

We treat all sub-bursts as individual bursts and use their features for machine learning algorithms. To provide a more comprehensive description of an FRB event, we included a broad range of parameters based on their usage in previous research, as different studies emphasize different features and identify varying important ones \citep[e.g.][]{Chen2022, Luo:2022smj, Zhu-Ge:2022nkz, Sun:2024huw}. These parameters can be divided into two categories: observational parameters (original data provided by Cat1 and Cat2023) and derived parameters, and we present the distribution of all parameters for the entire FRB dataset in Figure \ref{fig:fd}. We chose 10 observational parameters as did by \cite{Chen2022}: 
\begin{itemize}
    \item Boxcar width \bcwidth\,(ms) -- The boxcar width of the burst, with the label name `\textit{bc\_width}' in the two catalogs.
    
    \item Width of the burst \widthfitb\,(ms) -- The width of the burst that is fitted by \texttt{fitburst}\,\footnote{\url{https://github.com/CHIMEFRB/fitburst}}, with the label name `\textit{width\_fitb}' in the two catalogs.
    
    \item Flux \flux\,(Jy) -- The peak flux of the band-average profile (lower limit) with the label name `\textit{flux}' in the two catalogs.
    
    \item Fluence \fluence\,($\rm Jy \cdot ms$) -- The flux integrated over the duration of the burst (lower limit) with the label name `\textit{fluence}' in the two catalogs. 
    
    \item Scattering time \scattime\,(ms) -- The scattering time at 600 MHz of the burst, with the label name `\textit{scat\_time}' in the two catalogs.
    
    \item Spectral index \spidx \ -- The spectral shape parameter of the burst. The label name in the two catalogs is `\textit{sp\_idx}'.
    
    \item Spectral running \sprun \ -- This value characterizes the frequency dependence of the spectral shape and is labeled as `\textit{sp\_run}' in both catalogs.

    \item Highest frequency \highfreq\,(MHz) -- The highest frequency band of detection for the burst at full-width tenth-maximum. The label name in the two catalogs is `\textit{high\_freq}'.
    
    \item Lowest frequency \lowfreq\,(MHz) -- The lowest frequency band of detection for the burst at full-width tenth-maximum. The label name in the two catalogs is `\textit{low\_freq}'.
    
    \item Peak frequency \peakfreq\,(MHz) -- The peak frequency for the burst and is labeled as `\textit{peak\_freq}' in both catalogs. 
    
\end{itemize}
For \bcwidth, \widthfitb, \flux, \fluence, and \scattime, we take their logarithmic values throughout this work. Cat1 and Cat2023 only provide upper limits for the width of the burst and scattering time for some FRBs, so we opted to use these upper limits in our analysis. 

For the derived parameters, we choose 6 physical properties of FRBs \citep[see more details in][]{zhuge2023}: 
\begin{itemize}
    \item Redshift $z$  -- The redshift of FRBs is numerically derived from their dispersion measure (DM).
    
    In astronomical observations, distance plays a crucial role in analyzing the origin of FRBs. The Dispersion Measure (DM), indicating the total column density of free electrons along the line of sight, serves as a key distance-related parameter. For most FRBs, DM values far exceed those predicted for the Milky Way, pointing to their extragalactic origins. The DM of FRB can be separated into different components \citep[see e.g.][]{Deng:2013aga, 
    Gao:2014iva,
    Zhou:2014yta,
    Yang:2017bls,
    Yang:2016zbm, Li:2019klc, Wei:2019uhh, Qiang:2019zrs,Qiang:2020vta,Qiang:2021bwb,Qiang:2021ljr}:
    \begin{equation}
        {\rm DM_{obs} = DM_{MW} + DM_{halo} + DM_{IGM} + DM_{host}}/(1+z),
    \end{equation}
    where $\rm DM_{MW}, DM_{halo}, DM_{IGM} $ and $\rm DM_{host}$ represent the contributions from the Milky Way, the Milky Way halo, the intergalactic medium (IGM), and the host galaxy (including the interstellar medium of the host galaxy and the plasma around source) of the FRB. In the literature, we usually use extragalactic DM for research 
    \begin{equation}\label{eq:dme}
         {\rm DM_{E} = DM_{obs} - DM_{MW} - DM_{halo} = DM_{IGM} + DM_{host}}/(1+z).
    \end{equation}
    In this paper, the values $\rm DM_{obs}$ are provided by both catalogs with label 
    name `\textit{bonsai\_dm}'. We used the values of $\rm DM_{MW}$ as provided in Cat1 
    and Cat2023, which were estimated using the NE2001 model \citep{ne2001}(the 
    corresponding label names in Cat1 and Cat2023 are `\textit{dm\_exc\_ne2001}' and 
    `\textit{dm\_exc\_1\_ne2001}', respectively). Following the previous studies, we 
    adopt ${\rm DM_{halo} }= 30$\dmunit and ${\rm DM_{host}} = 70$\dmunit. For the
    $\rm DM_{IGM}$, it can be written as \citep[see e.g.][]{Deng:2013aga, Yang:2016zbm, Li:2019klc, Wei:2019uhh, Qiang:2019zrs,Qiang:2020vta,Qiang:2021bwb,Qiang:2021ljr}
    \begin{equation}\label{eq:dmigm}
        {\rm DM_{IGM}}(z)=\frac{3cH_0\Omega_{b,0}}{8\pi G m_p}\int_0^z\frac{f_{\rm IGM}(\tilde{z})\,f_e(\tilde{z})\left(1+\tilde{z}\right)d\tilde{z}}{E(\tilde{z})},
    \end{equation}
    where $c$ is the speed of light, $G$ is the gravitational constant, $m_p$ is the mass of proton, $E(z)$ is the dimensionless Hubble parameter. For cosmological parameters, we adopt $H_{0}=67.4$\hunit, $\Omega_{m}=0.315$ and $\Omega_{b,0}h^{2}=0.0224$ from the latest Planck 18 results for the flat $\Lambda$CDM cosmology \citep{Planck:2018vyg}. 
    $f_e(z)$ is the ionized electron number fraction per baryon, and $f_{\rm IGM}(z)$ is the fraction of baryon mass in IGM. In principle, these two parameters are functions of redshift z. In this work, we follow e.g. \cite{Qiang:2019zrs, Qiang:2021bwb, Gao:2014iva, Yang:2017bls} and use the fiducial values $f_e = 0.875$ and $f_{\rm IGM} = 0.83$. According to Eq.\ref{eq:dme} and Eq. \ref{eq:dmigm}, we can derive the redshift of all FRBs. Following \cite{zhuge2023}, we also set a minimum redshift of 0.002248 corresponding to a luminosity distance of 10 Mpc to avoid zero or negative values.
    
    \item Rest-frame frequency width $\Delta \nu$\,(MHz) -- The frequency width corrected for the cosmological redshift effect, it can be calculated by
    \begin{equation}
        \Delta \nu = (\nu_{max}-\nu_{min})(1+z).
    \end{equation}
    
    \item Rest-frame width \rw\,(ms) -- The width of the burst \widthfitb \ corrected for the cosmological redshift effect: \rw \,=\,\widthfitb$/(1+z)$. We take the logarithmic values.
    
    \item Burst energy $E$\,(erg) -- The energy of a FRB can be calculated by 
    \begin{equation}
        E = \frac{4\pi d_{L}^{2}}{1+z} F_{\nu} \nu_{p},
    \end{equation}
    where $d_L$ is the luminosity distance. We take their logarithmic values.
    
    \item Luminosity $L$\,(erg/s) -- The luminosity of FRBs can be derived from \cite{}
    \begin{equation}
        L = 4\pi d_{L}^{2} S_{\nu} \nu_{p},
    \end{equation}
    and we take the logarithmic values.
    
    \item Brightness temperature \bt\,(K) -- The brightness temperature can be derived as \citep{Luo:2022smj}
    \begin{equation}
        T_{B} = \frac{S_{\nu}d_{L}^{2}}{2\pi \kappa_{B}(\nu_{p}\Delta t_{fitb})^{2}} = 1.1 \times 10^{35} \, {\rm K} \, \left(\frac{S_{\nu}}{\rm Jy}\right)\, \left(\frac{d_{L}}{\rm Gpc}\right)^{2}\, \left(\frac{\nu_{p}}{\rm GHz}\right)^{-2}\, \left(\frac{\Delta t_{fitb}}{\rm ms}\right)^{-2} \, \frac{1}{1+z}
        ,
    \end{equation}
    
    where $\kappa_{B}$ is the Boltzmann constant. We take the logarithmic values.
\end{itemize}

\section{Method} \label{sec:met}
Dimensionality reduction and clustering are two types of unsupervised machine learning methods used in this paper. First, we apply the dimensionality reduction algorithm to automatically convert high-dimensional data into low-dimensional data. Then, we use a clustering algorithm to cluster the reduced-dimensional data based on their similarities.

\subsection{Machine Learning Techniques}

\subsubsection{Dimensionality Reduction}\label{subsec:dr}
In this study, we use Uniform Manifold Approximation and Projection \citep[UMAP,][]{umap}, implemented via the Python package \texttt{umap-learn}\,\footnote{\url{https://github.com/lmcinnes/umap}}, to perform dimensionality reduction. UMAP is a dimensionality reduction technique that can be used for both visualization and general nonlinear dimensionality reduction. This algorithm assumes that the input data is uniformly distributed on a Riemannian manifold with a locally constant (or approximately constant) Riemannian metric and that the manifold is locally connected. Based on these assumptions, the manifold can be modeled using a fuzzy topological structure.

The use of UMAP has been extensively explored in many studies \citep[e.g.][]{Zhu-Ge:2022nkz, Yang:2023dcf, Raquel:2023tgi, Chen:2023pub}. For FRB classification, the three parameters \texttt{n\_components}, \texttt{n\_neighbors}, and \texttt{min\_dist} have a more significant impact on the classification results. Meanwhile, we also experimented with modifying other parameters and found that they had minimal effect on the classification outcomes. Therefore, in this study, we choose to focus on adjusting these three parameters of UMAP. \texttt{n\_components} allows us to determine the dimensionality of the reduced space where the data will be embedded. In our work, we set $\mathtt{n\_components}=2$ for all features, projecting the data onto a 2D plane for visual representation. \texttt{n\_neighbors} controls how UMAP balances the local and global structure of data. UMAP achieves this by controlling the size of the local neighborhood it considers when attempting to learn the underlying structure of the data. This means that low values of \texttt{n\_neighbors} will cause UMAP to focus on very local structures, potentially at the cost of missing the overall global structure. On the other hand, higher values of \texttt{n\_neighbors} will push UMAP to consider larger neighborhoods around each point, capturing the broader structure of the data, but possibly losing finer details. \texttt{min\_dist} decides how tightly UMAP can pack points together in low-dimensional space. Basically, it sets the minimum distance between points in the low-dimensional space. Lower values of \texttt{min\_dist} will lead to more tightly packed, ``clumpier" embeddings, which can be beneficial for identifying clusters or preserving finer topological details. In contrast, higher values of \texttt{min\_dist} will prevent the points from being tightly packed, instead focusing on maintaining the broader topological structure. We scan \texttt{n\_neighbors} from 2 to 50 and \texttt{min\_dist} from 0.0 to 0.99. In this paper, we take \texttt{n\_neighbors} = 21, and \texttt{min\_dist} = 0.03.

%===============fig:sch================
\begin{figure}
    \centering
    \includegraphics[width=1\linewidth]{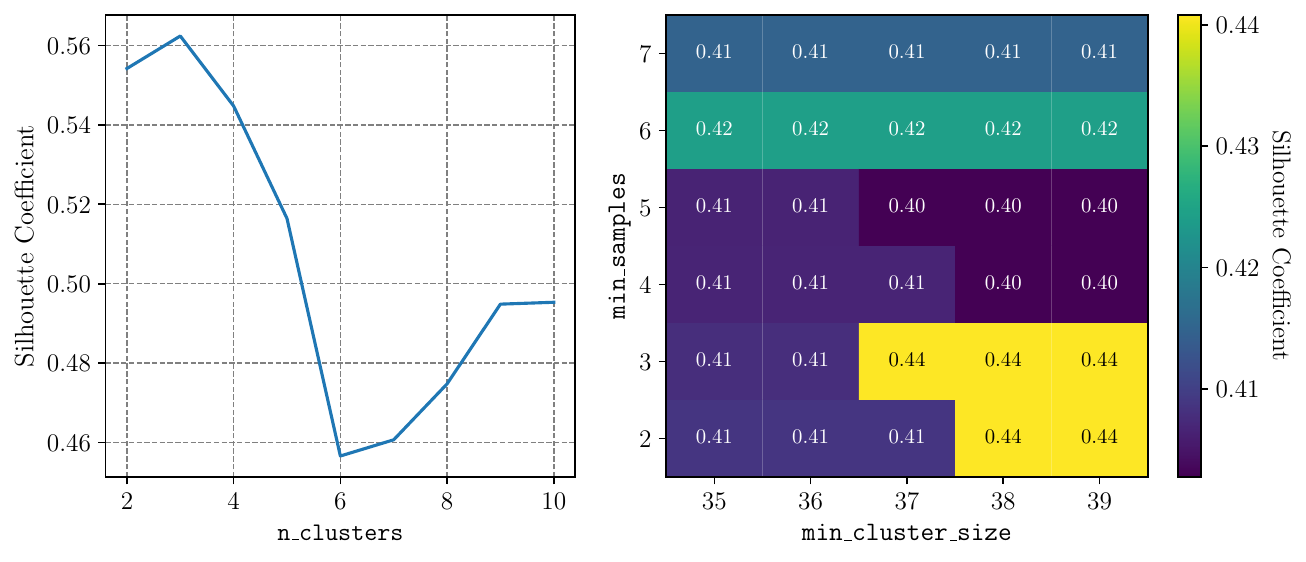}
    \caption{\label{fig:sch} The mean silhouette coefficients of the k-means (left panel) and HDBSCAN (right panel) with respect to different hyperparameters.}
\end{figure}

\subsubsection{Clustering Algorithms}\label{subsec:cls}
In this work, we used k-means \citep{macqueen1967_k-means,lloyd1982_k-means} and Hierarchical Density-Based Spatial Clustering of Applications with Noise \citep[HDBSCAN,][]{Campello_2013_HDBSCAN,Campello2015-HDBSCAN,McInnes2017_HDBSCAN} to group the reduced-dimensional data into different clusters. K-means clustering is based on the distance between each data point and its corresponding cluster center. It aims to minimize the distance between the points and their respective centers, effectively grouping similar points into clusters based on proximity in the feature space. Initially, the k-means algorithm selects k random points as the initial cluster centers, calculates the Euclidean distance of each data point from these centers, and assigns each point to the nearest center. Then, it recalculates the mean of each cluster, updating the cluster centers. This process is repeated iteratively until the cluster centers stabilize, minimizing the overall variance within the cluster \citep[for details, see][and references therein]{umlrevi}. We use \texttt{sklearn.cluster.KMeans}\,\footnote{\url{https://scikit-learn.org/stable/modules/generated/sklearn.cluster.KMeans.html}} to perform the k-means clustering algorithm. The essential hyperparameter is \texttt{n\_clusters} which means how many clusters are present in the model.

HDBSCAN is a clustering algorithm developed by \cite{Campello_2013_HDBSCAN, Campello2015-HDBSCAN}. It extends Density-based Spatial Clustering of Applications with Noise \citep[DBSCAN,][]{ester1996_dbscan} by transforming it into a hierarchical clustering method \citep{HAN2012443}. The algorithm then extracts flat clusters from this hierarchy based on the stability of the clusters, which allows it to better handle varying densities and identify clusters of different shapes and sizes. HDBSCAN uses minimum spanning trees, allowing it to discover clusters with varying densities, unlike DBSCAN, which assumes a constant density across the entire dataset. This flexibility enables HDBSCAN to identify clusters of different shapes and sizes, making it more adaptable to complex structures. We use a Python package \texttt{hdbscan}\,\footnote{\url{https://hdbscan.readthedocs.io/en/latest/index.html}} to perform this clustering algorithm. The hyperparameters of HDBSCAN adjusted in this paper are \texttt{min\_cluster\_size} and \texttt{min\_samples}. \texttt{min\_cluster\_size} controls the minimum number of points required to form a cluster, while \texttt{min\_samples} determines how conservative the algorithm is in classifying points as noise or part of a cluster. These parameters directly influence the number of clusters and the overall shape of the clustering.

 To improve the clustering results, we optimize the hyperparameters of the clustering algorithm by maximizing the mean silhouette coefficient\,\footnote{\url{https://scikit-learn.org/stable/modules/generated/sklearn.metrics.silhouette_score.html}} \citep{silhouettes}  across all samples. The silhouette coefficient measures how well a sample fits within its assigned cluster, with values ranging from -1 to 1. Higher values indicate more tightly grouped, well-defined clusters. We show the mean silhouette coefficients changed with the hyperparameters of the clustering algorithm in Figure \ref{fig:sch}. In this work, we chose the hyperparameters with max mean silhouette coefficients, which are \texttt{n\_clusters} = 3 for k-means, \texttt{min\_cluster\_size} = 37\,\footnote{In fact, when we fixed \texttt{min\_samples} at 3, the mean silhouette coefficients remained constant as \texttt{min\_cluster\_size} varied from 37 to 60. However, the number of noise points increased during this range. Therefore, we chose \texttt{min\_cluster\_size} = 37 for this study.} and \texttt{min\_samples} = 3 for HDBSCAN. We also include the condensed tree of HDBSCAN in Appendix \ref{app:hct} to highlight the persistence and stability of the clustering results.

\subsection{Evaluation Metrics}\label{subsec:em}

In this study, we experimented with various machine learning algorithms and hyperparameters, and their classification performance requires evaluation using specific metrics. The outputs of clustering can be written as the following forms:
\begin{itemize}
    \item $TP$: The true positives, which means the number of repeaters correctly classified in the repeater cluster.
    \item $TN$: The true negatives, represent the number of non-repeaters correctly classified in the non-repeater cluster.
    \item $FP$: The false positives, represent the number of non-repeaters incorrectly classified in the repeater cluster.
    \item $FN$: The false negatives, indicating the number of repeaters incorrectly classified into the non-repeater cluster.
\end{itemize}
Generally, based on the four outputs mentioned above, various metrics can be calculated to evaluate the model's performance:
\begin{itemize}
    \item Recall: ${TP}/{(TP+FN)}$.
    \item Precision: ${TP}/{(TP+FP)}$.
    \item Accuracy: ${(TP+TN)}/{(TP+TN+FP+FN)}$.
\end{itemize}
In this study, we use recall to evaluate the model's performance, as observational limitations prevent the accurate determination of non-repeaters, making it impossible to reliably estimate $TN$ and $FP$.

\section{Results and Discussion} \label{sec:res}
\subsection{Dimensionality reduction and clustering}\label{subsec:res1}

\begin{figure}
    \centering
    \includegraphics[width=1\linewidth]{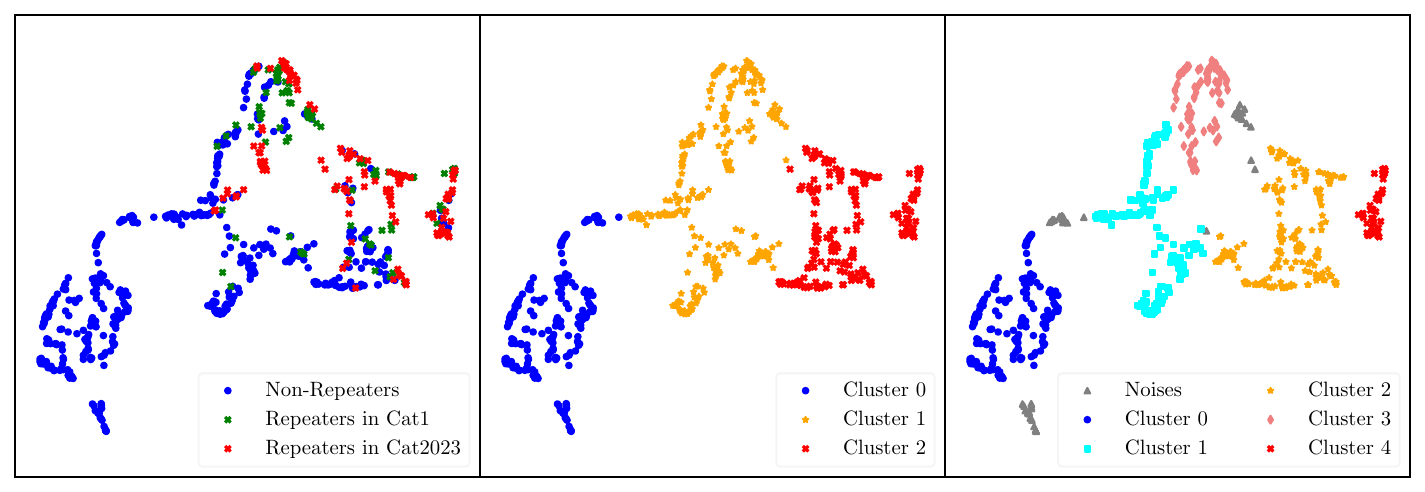}
    \caption{\label{fig:ukh}The results of  dimensionality reduction and clustering. Left panel: The distribution of UMAP-embedded data. Blue dots represent non-repeaters, while green and red crosses indicate repeaters from Cat1 and Cat2023, respectively. Middle panel: The clustering result of UMAP-embedded data from k-means. Cluster 0, 1, and 2 are represented by blue dots, orange stars, and red crosses, respectively. Right panel: The clustering result of UMAP-embedded data from HDBSCAN. Gray triangles, blue dots, cyan squares, orange stars, light coral diamonds, and red crosses represent Noise, and Cluster 0, 1, 2, 3, and 4, respectively.}
\end{figure}

\begin{table}[]
\centering
\renewcommand{\arraystretch}{1.5}  % 控制表格行高
\begin{tabular}{|c|c|c|c|c|c|c|c|c|}
\hline
Algorithm & Clusters &\makecell[c]{\hspace{-1.2cm}Non-repeater\\ \hspace{-1.2cm}number } &\makecell[c]{\hspace{-1.2cm}Repeater\\ \hspace{-1.2cm}number} &\makecell[c]{\hspace{-1.2cm}Repeater\\ \hspace{-1.2cm}candidate\\ \hspace{-1.2cm}number} &\makecell[c]{\hspace{-1.2cm}Total\\ \hspace{-1.2cm}number} &\makecell[c]{\hspace{-1.2cm}Repeater\\ \hspace{-1.2cm}burst\\ \hspace{-1.2cm}percentage} & Recall&\makecell[c]{\hspace{-1.2cm}Repeater\\ \hspace{-1.2cm}source\\ \hspace{-1.2cm}percentage} \\ \hline
\multirow{3}{*}{K-means} & 0 & \centering 210 & 0 & 0 & 210 & 0\% & \multirow{3}{*}{100\%}  & \multirow{3}{*}{61.7\%} \\ \cline{2-7}
                         & 1 & 0 & 98 & 196 & 294 & 33.3\% & & \\ \cline{2-7}
                         & 2 & 0 & 132 & 103 & 235 & 56.2\% & & \\ \hline
\multirow{6}{*}{HDBSCAN} & Noise     & 45 & 14 & 0 & 59 & 23.7\% & \multirow{6}{*}{93.5\%} & \multirow{6}{*}{37.0\%} \\ \cline{2-7}
                         & 0 & 169 & 0 & 0 & 169 & 0\% & & \\ \cline{2-7}
                         & 1 & 138 & 15 & 0 & 153 & 9.8\% & & \\ \cline{2-7}
                         & 2 & 0 & 88 & 118 & 206 & 42.7\% & & \\ \cline{2-7}
                         & 3 & 0 & 66 & 28 & 94 & 70.2\% & & \\ \cline{2-7}
                         & 4 & 0 & 47 & 11 & 58 & 81.0\% & & \\ \hline
\end{tabular}
\vspace{5mm}
\caption{\label{tab:cls}The clustering results of UMAP-embedded data using k-means and HDBSCAN. For a detailed description, see Section \ref{subsec:res1}.}
\end{table}

Based on the methods and hyperparameters discussed in Section \ref{subsec:dr}, we present the UMAP-dimensionally reduced features of 745 FRBs in the left panel of Figure \ref{fig:ukh}. The blue dots represent non-repeaters, while the green and red crosses indicate repeaters from Cat1 and Cat2023, respectively. It is evident that after dimensionality reduction by UMAP, the repeaters are clustered in the upper right corner, while a distinct group of pure non-repeaters appears in the lower left corner, separated by a noticeable gap from the mixture. This indicates that the 16 parameters of FRBs used in the analysis have the ability to distinguish between non-repeaters and repeaters.

\begin{figure}
    \centering
    \includegraphics[width=1\linewidth]{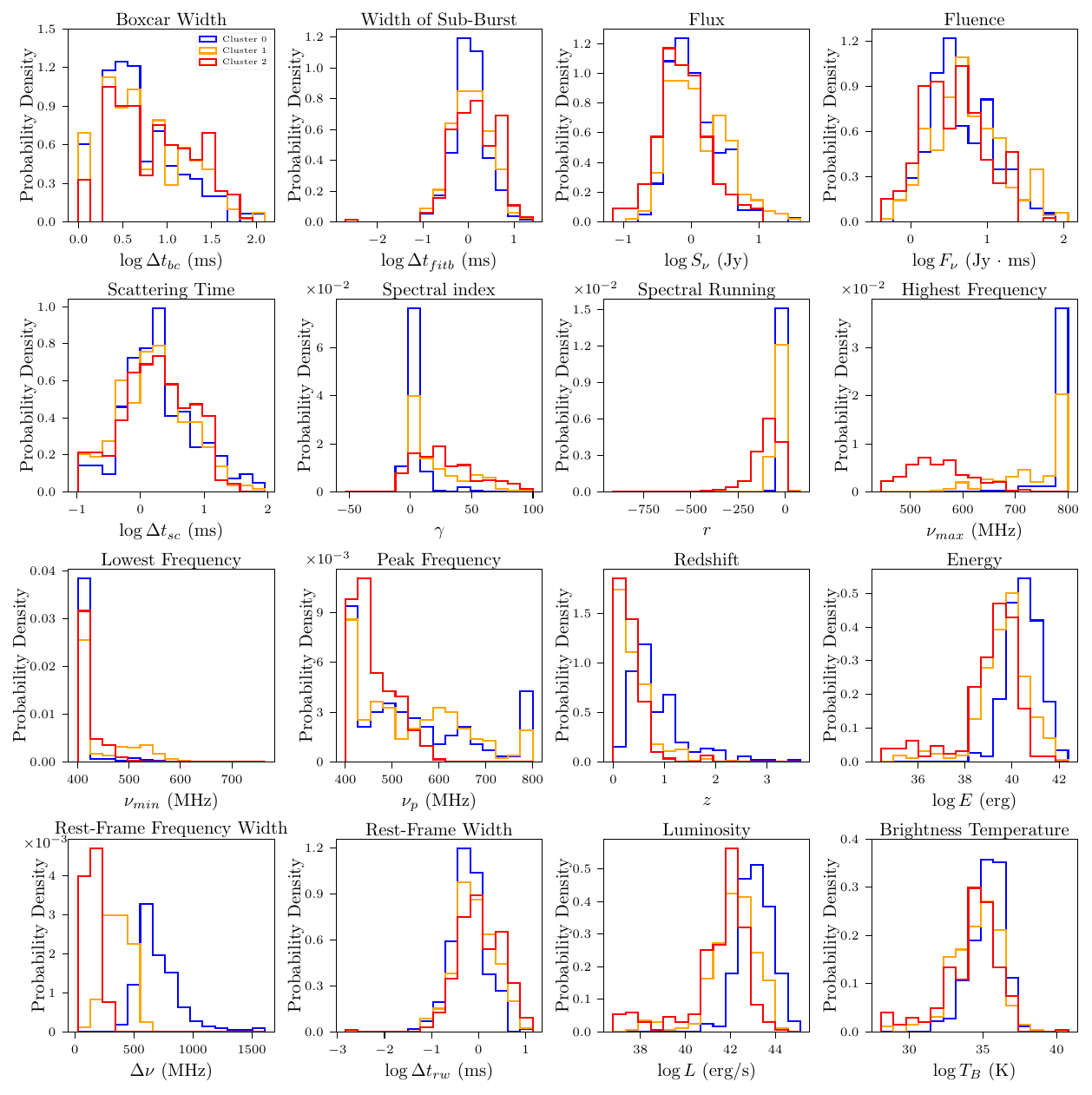}
    \caption{\label{fig:kcfd}The distributions of 16 parameters across different clusters from k-means are illustrated, with blue, orange, and red histogram step lines representing Cluster 0, Cluster 1, and Cluster 2, respectively.}
    
\end{figure}

\begin{figure}
    \centering
    \includegraphics[width=1\linewidth]{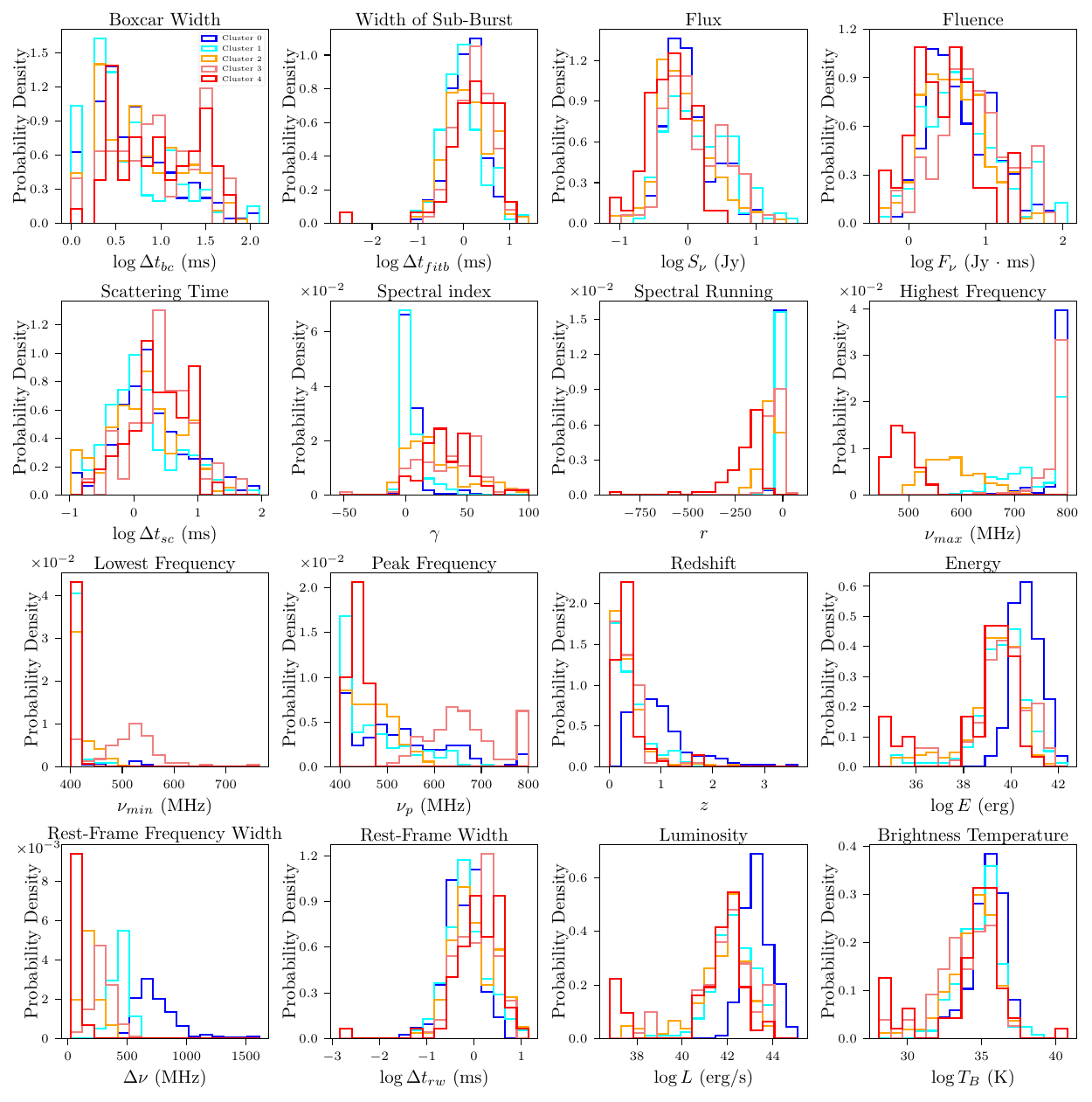}
    \caption{\label{fig:hcfd}Same as Figure \ref{fig:kcfd}, but displays the results from HDBSCAN. The blue, cyan, orange, light coral, and red histogram step lines represent Cluster 0, 1, 2, 3, and 4, respectively.}
    
\end{figure}
Then we used two types of clustering algorithms, k-means and HDBSCAN, to cluster the two-dimensional UMAP embedding. We present the clustering results in the middle and right panels of Figure \ref{fig:ukh}, based on the hyperparameters of k-means and HDBSCAN discussed in Section \ref{subsec:cls}. If the proportion of repeaters in a cluster exceeds 30\%, we classify it as a `` repeater cluster ", with the non-repeaters within it considered as `` repeater candidates ". Conversely, if the proportion is below 30\%, it is classified as an `` non-repeater cluster ". 

As shown in the middle panel of Figure \ref{fig:ukh}, the k-means algorithm divided the UMAP-embedded data into three clusters. Cluster 0 contains 210 FRBs, all of which are non-repeaters. Cluster 2 consists of 294 FRBs, including 98 repeaters and 196 repeater candidates (the repeater burst percentage is 33.3\%). Cluster 3 contains 235 FRBs, with 132 repeaters (the repeater burst percentage is 56.2\%) and 103 repeater candidates. All repeaters are classified into repeater clusters, giving a recall of 100\%. Out of the 509 non-repeaters, 299 repeater candidates were identified (from 269 non-repeater sources). If these repeater candidates are real, the repeater source percentage of FRBs would reach approximately 61.7\%, exceeding the predicted rates from studies such as \cite{Chen2022}, \cite{Zhu-Ge:2022nkz}, and \cite{Yang:2023dcf}.

As shown in the right panel of Figure \ref{fig:ukh}, HDBSCAN divided the UMAP-embedded data into five clusters and a noise cluster. Cluster 0 contains 169 FRBs, all of which are non-repeaters. Cluster 1 includes 15 repeaters and 138 non-repeaters, with repeater bursts making up only 9.8\%. Clusters 2, 3, and 4 contain 206, 94, and 58 FRBs, with 88, 66, and 47 repeaters (repeater bursts percentage accounting for 42.7\%, 70.2\%, and 81.0\%, respectively) and repeater candidates numbering 118, 28, and 11 (a total of 157 candidates, corresponding to 141 non-repeater sources). The recall for the 739 samples is 93.1\%. The overall repeater source percentage is around 37.9\%, slightly lower than the results of \cite{Yamasaki:2023dlb, McGregor:2023mzr}, but comparable to those of \cite{Chen2022}, \cite{Zhu-Ge:2022nkz}, and \cite{Yang:2023dcf}. Detailed clustering results for both algorithms are also presented in Table \ref{tab:cls}.

We plot the feature distributions of different clusters generated by k-means and HDBSCAN in Figure \ref{fig:kcfd} and Figure \ref{fig:hcfd}, respectively. As shown in Figure \ref{fig:kcfd}, the rest-frame frequency width differs the most across clusters, indicating that repeater clusters tend to have narrower frequency bandwidths. Significant differences are also observed in the distributions of the spectral index, highest frequency, redshift, energy, luminosity, and brightness temperature among the clusters. These results are consistent with the feature distribution of repeaters and non-repeaters shown in Figure \ref{fig:fd}. In Figure \ref{fig:hcfd}, almost all features show notable differences in their distributions among clusters, with rest-frame frequency width once again being the most distinct, similar to the results in Figure \ref{fig:kcfd}.

\begin{figure}
    \centering
    \includegraphics[width=0.98\linewidth]{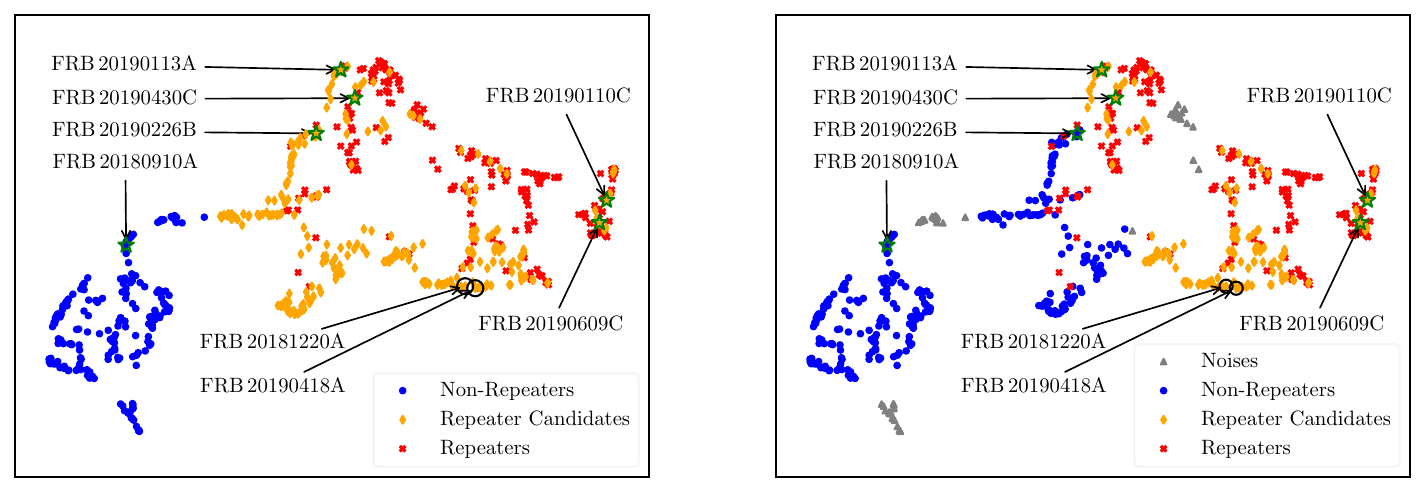}
    \caption{\label{fig:khpc}The distribution of non-repeaters (blue dots), repeater candidates (orange diamonds), and repeaters (red crosses) is presented in the clustered UMAP-embedded data from both k-means (left panel) and HDBSCAN (right panel). The gray triangles represent noises. The green stars mark the first bursts of six repeaters that were misidentified as non-repeaters in Cat1 but are later identified as repeaters in Cat2023. The black circles indicate the localized non-repeaters.}
    
\end{figure}

As mentioned in Section \ref{sec:data}, there are 6 FRB sources that were previously observed as non-repeaters in Cat1 but were later identified as repeaters in Cat2023. We marked these 6 FRBs in Figure \ref{fig:khpc} with green open stars to analyze the reliability of our repeater candidate predictions. The left and right panels of Figure \ref{fig:khpc} show the repeater candidate predictions from k-means and HDBSCAN, respectively. For k-means, five out  of these 6 FRB sources are located in the repeater clusters and were successfully identified as repeater candidates, while HDBSCAN predicted four of them successfully. This indicates that our method is effective in identifying potential repeaters. However, one of these 6 FRBs stands out—FRB\,20180910A—which was classified into the non-repeater cluster by both clustering algorithms. After analyzing its 16 parameters, we found that its boxcar width is only 0.98 ms, and its spectral index (0.05) and spectral running (-0.53) are very similar to those of non-repeaters. Additionally, its broadband emission characteristics differ significantly from the narrow-band emission typically observed in repeaters. Currently, there is no satisfactory theoretical explanation for repeaters that exhibit characteristics so similar to non-repeaters. We also found that FRB\,20180910A has so far produced three detected bursts, with intervals of 1 year and 9 months. These bursts exhibit noticeable differences in boxcar width, bandwidth, spectral index, and spectral running. It is possible that these bursts are from different non-repeaters within the same galaxy (or neighboring galaxies in the same direction). Further observations are needed to confirm this. 
\begin{figure}[ht]
    \centering
    \includegraphics[width=1\linewidth]{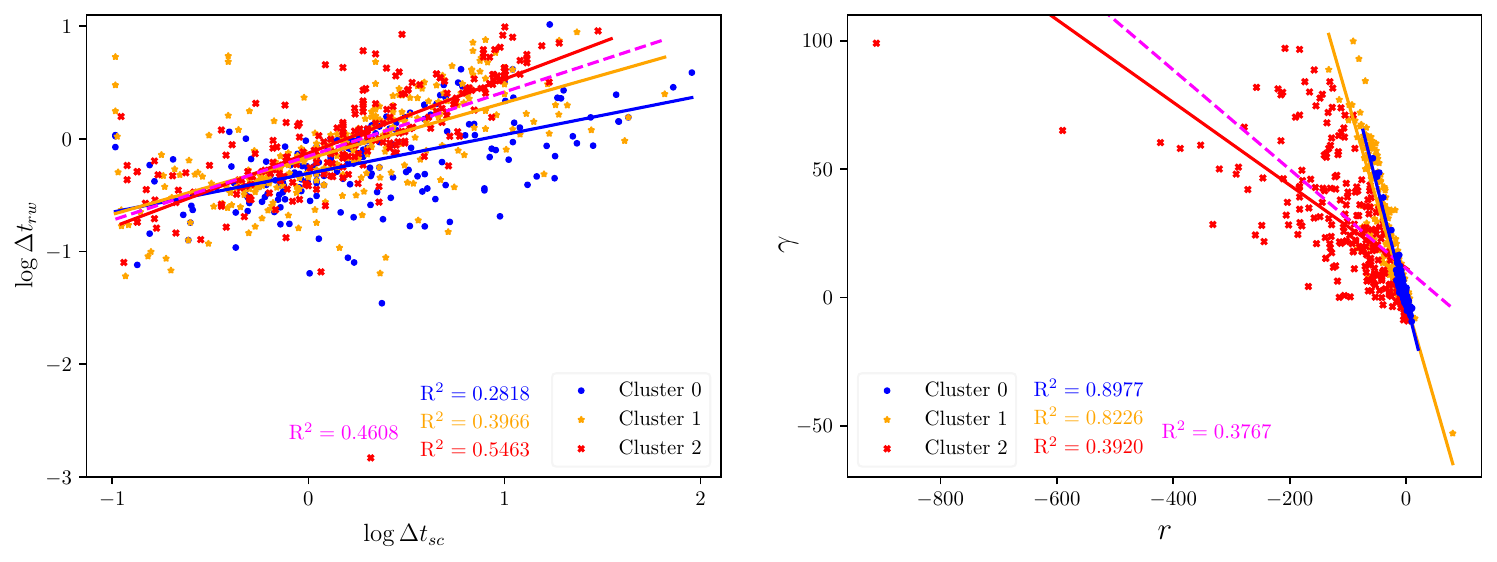}
    \caption{\label{fig:ker}The empirical relationships between scattering time (${\rm log \,}\Delta t_{sc}$) and rest-frame width (${\rm log \,}\Delta t_{rw}$) (left panel), and between spectral acceleration ($r$) and spectral index ($\gamma$) (right panel) are shown for FRBs in different clusters from k-means. These relationships are represented by blue, orange, and red lines, corresponding to clusters 0, 1, and 2, respectively. The associated $R^2$ values are shown in the same colors. Additionally, the dashed magenta line and its corresponding $R^2$ represent the combined data from clusters 1 and 2. See Section \ref{subsec:res2} for details.}

\end{figure}
\begin{table}[ht]
\centering
\renewcommand{\arraystretch}{1.2}  % 控制表格行高
\setlength{\tabcolsep}{26pt}
\begin{tabular}{|c|c|c|c|c|c|}
\hline
x                                            & y                                            & Clusters & $a$       & $b$       & $R^2$ \\ \hline
\multirow{4}{*}{${\rm log \,}\Delta t_{sc}$} & \multirow{4}{*}{${\rm log \,}\Delta t_{rw}$} & 0       & 0.3440 & -0.3055  & 0.2818               \\ \cline{3-6} 
                                             &                                              & 1       & 0.4954 & -0.1744  & 0.3966               \\ \cline{3-6} 
                                             &                                              & 2       & 0.6572 & -0.1256  & 0.5463               \\ \cline{3-6} 
                                             &                                              & [\,1, 2\,]     & 0.5691 & -0.1519  & 0.4608               \\ \hline
\multirow{4}{*}{$r$}                           & \multirow{4}{*}{$\gamma$}                  & 0       & -0.9000 & -1.6846 & 0.8977               \\ \cline{3-6} 
                                             &                                              & 1       & -0.7853 & -1.9914 & 0.8226               \\ \cline{3-6} 
                                             &                                              & 2       & -0.1613 & 11.4422 & 0.3920               \\ \cline{3-6} 
                                             &                                              & [\,1, 2\,]     & -0.1936 & 11.1439 & 0.3767               \\ \hline

\end{tabular}
\caption{\label{tab:ker}The slope ($a$), intercept ($b$), and $R^2$ values of the empirical relations  for different clusters from k-means.
}
\end{table}
\begin{table}[]
\centering
\begin{tabular}{|c|c|c|}
\hline
\diagbox[width=5cm]{Clusters}{$p$-value}{Relation} & ${\rm log \,}\Delta t_{sc} 
\,-\,{\rm log \,}\Delta t_{rw}$ & $r \,-\, \gamma$ \\ \hline
0 \& 1            & 0.0016                                                   & 0.5177        \\ \hline
0 \& 2            & 1.75e-06                                                 & 6.28e-05      \\ \hline
1 \& 2            & 0.1975                                                   & 1.11e-16      \\ \hline
0 \& {[}1, 2{]}   & 2.49e-06                                                 & 0.0190        \\ \hline
\end{tabular}
\caption{\label{tab:kct} The $p$-values of Chow tests for k-means clusters.
}
\end{table}
\hspace{-4mm}Regarding FRB\,20190226B, which was misidentified only by HDBSCAN, we analyzed its 16 parameters and found no distinct characteristics suggesting it is a non-repeating burst.

Additionally, some of the non-repeaters from Cat1 have identified host galaxies \citep{Bhardwaj2024ApJ, Law2020ApJ}. In both panels of Figure \ref{fig:khpc}, we highlight two of these FRBs (FRB\,20181220A and FRB\,20190418A) that were identified as repeater candidates by both clustering algorithms with black circles. Continued observations of the host galaxies of these two FRBs may reveal further repeating bursts in the future.

\begin{figure}
    \centering
    \includegraphics[width=1\linewidth]{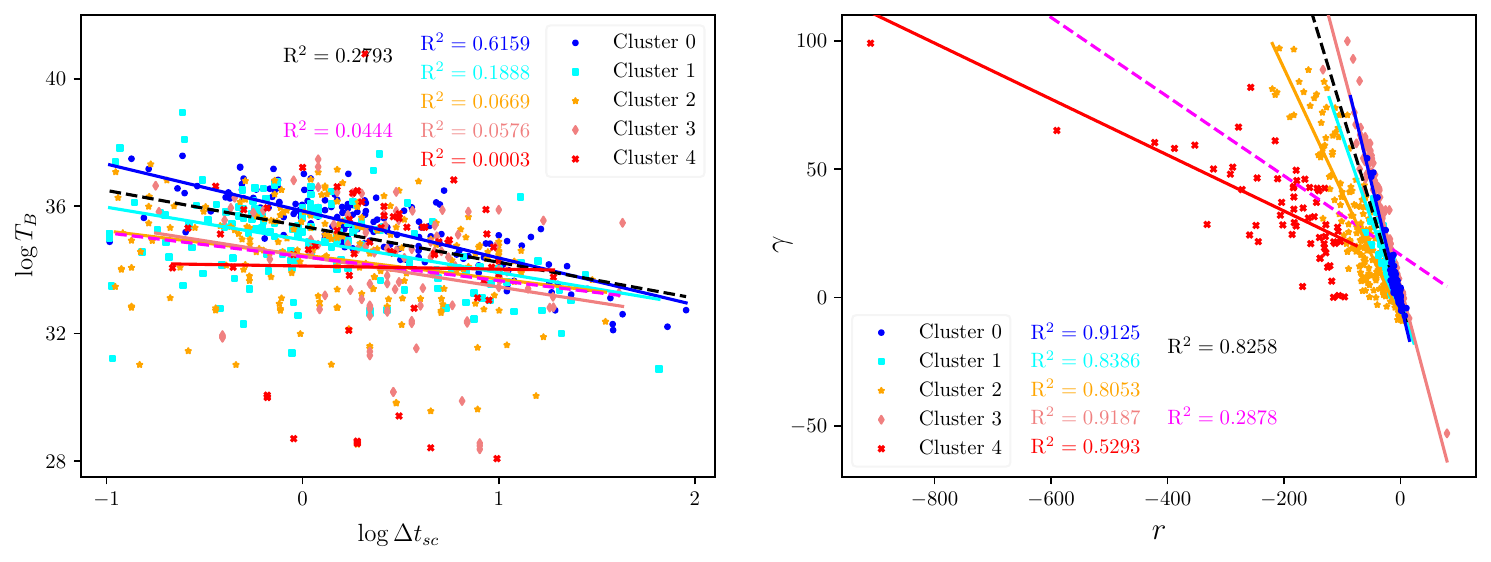}
    \caption{\label{fig:her}The empirical relationships between scattering time (${\rm log \,}\Delta t_{sc}$) and brightness temperature (${\rm log \,}T_B$) (left panel), and between spectral acceleration ($r$) and spectral index ($\gamma$) (right panel) are shown for FRBs in different clusters from HDBSCAN. These relationships are represented by blue, cyan, orange, light coral, and red lines, corresponding to clusters 0, 1, 2, 3, and 4, respectively. The associated $R^2$ values are shown in the same colors. Additionally, the dashed black (magenta) line and its corresponding $R^2$ represent the combined data from clusters 0 and 1 (2, 3, and 4). See Section \ref{subsec:res2} for details.}
\end{figure}

\begin{table}[ht]
\centering
\renewcommand{\arraystretch}{1.2}  % 控制表格行高
\setlength{\tabcolsep}{28pt}
\begin{tabular}{|c|c|c|c|c|c|c}
 \cline{1-6}
x                                            & y                                    & Clusters & $a$     & $b$     & $R^2$  &  \\ \cline{1-6}
\multirow{7}{*}{${\rm log \,}\Delta t_{sc}$} & \multirow{7}{*}{${\rm log \,}T_{B}$} & 0       & -1.4741 & 35.8475 & 0.6159 &  \\ \cline{3-6}
                                             &                                      & 1       & -1.0264 & 34.9418 & 0.1888 &  \\ \cline{3-6}
                                             &                                      & 2       & -0.7559 & 34.4684 & 0.0669 &  \\ \cline{3-6}
                                             &                                      & 3       & -0.9667 & 34.4318 & 0.0576 &  \\ \cline{3-6}
                                             &                                      & 4       & -0.098  & 34.1223 & 0.0003 &  \\ \cline{3-6}
                                             &                                      & [\,0, 1\,]   & -1.1274 & 35.3659 & 0.2793 &  \\ \cline{3-6}
                                             &                                      & [\,2, 3, 4\,] & -0.7510 & 34.4152 & 0.0444 &  \\ \cline{1-6}
\multirow{7}{*}{$r$}                         & \multirow{7}{*}{$\gamma$}            & 0       & -0.9394 & -2.2495 & 0.9125 &  \\ \cline{3-6}
                                             &                                      & 1       & -0.6574 & -2.6236 & 0.8386 &  \\ \cline{3-6}
                                             &                                      & 2       & -0.4835 & -7.4780 & 0.8053 &  \\ \cline{3-6}
                                             &                                      & 3       & -0.8527 & 4.4482  & 0.9187 &  \\ \cline{3-6}
                                             &                                      & 4       & -0.1090 & 11.8615 & 0.5293 &  \\ \cline{3-6}
                                             &                                      & [\,0, 1\,]   & -0.7424 & -2.1286 & 0.8258 &  \\ \cline{3-6}
                                             &                                      & [\,2, 3, 4\,] & -0.1542 & 16.5526 & 0.2878 &  \\ \cline{1-6}
\end{tabular}
 \caption{\label{tab:her}The slope ($a$), intercept ($b$), and $R^2$ values of the empirical relations  for different clusters from HDBSCAN.}
\end{table}
\begin{table}[]
\centering
\begin{tabular}{|c|c|c|}
\hline
\diagbox[width=5cm]{Clusters}{$p$-value}{Relation} & ${\rm log \,}\Delta t_{sc} 
\,-\,{\rm log \,}T_B$ & $r \,-\, \gamma$ \\ \hline
0 \& 1            & 0.0041                                                   & 1.13e-10        \\ \hline
0 \& 2            & 0.1701                                                 & 1.75e-09      \\ \hline
0 \& 3            & 0.0483                                                 & 0.4561      \\ \hline
0 \& 4            & 0.5232                                                 & 1.11e-16      \\ \hline
1 \& 2            & 0.2403                                                   & 0.0019      \\ \hline
1 \& 3            & 8.12e-07                                                   & 6.34e-12      \\ \hline
1 \& 4            & 0.0014                                                   & 1.11e-16      \\ \hline
2 \& 3            & 0.0002                                                   & 1.11e-16      \\ \hline
2 \& 4            & 0.0379                                                   & 1.11e-16      \\ \hline
3 \& 4            & 0.6024                                                   & 1.11e-16      \\ \hline
{[}0, 1{]} \& {[}2, 3, 4{]}   & 2.43e-11                                     & 1.97e-08        \\ \hline
\end{tabular}
\caption{\label{tab:hct} The $p$-values of Chow tests for HDBSCAN clusters.
}
\end{table}
\subsection{Empirical Relationships}\label{subsec:res2}
In this paper, we further analyze the potential two-dimensional empirical relations within different clusters identified by various clustering algorithms. We pair the 16 parameters of FRBs from different clusters and linearly fit the data points using \texttt{scipy.stats.linregress}\,\footnote{\url{https://docs.scipy.org/doc/scipy/reference/generated/scipy.stats.linregress.html}}, which performs linear least-squares regression. The form of the two-dimensional empirical relationship is given as $y = a\,x + b$, and the goodness of fit is evaluated using the score (coefficient of determination), defined as $R^2=1-\sum_{i}(y_i-\hat{y}_i)^2/\sum_{i}(y_i-\bar{y}_i)^2$, where $y_i$, $\hat{y}_i$, and $\bar{y}_i$ are the observed values, the regressed values and the mean of the observed values, respectively. The closer $R^2$ is to 1,  the better the model fits the data. To quantify the statistical differences in the slopes and intercepts of empirical relationships across different clusters, we implemented the Chow test for evaluation \citep{chowtest}. The null hypothesis of the Chow test is that the two samples come from the same regression model. If the $p$-value is less than 0.05, the null hypothesis is rejected, indicating that the two samples come from different regression models.

We set a high score threshold of $R^2>0.5$ to filter out well-fitted empirical relations and excluded parameter combinations that inherently have a linear relationship (e.g., luminosity and flux). For the clusters from k-means, the selected empirical relations are shown in Figure \ref{fig:ker}. The slope, intercept, and score of the empirical relations are listed in Table \ref{tab:ker}. The $p$-values of the Chow test between different clusters are listed in Table \ref{tab:kct}. The left panel of Figure \ref{fig:ker} shows the empirical relation between two independent parameters: scattering time (${\rm log \,}\Delta t_{sc}$) and rest-frame width (${\rm log \,}\Delta t_{rw}$). We observe that $R^2>0.5$ only in cluster 2, corresponding to a repeater cluster, whereas the ${\rm log \,}\Delta t_{sc} - {\rm log \,}\Delta t_{rw}$ relation appears less significant in the other two clusters. However, the slope and intercept of the relation ${\rm log \,}\Delta t_{sc} - {\rm log \,}\Delta t_{rw}$ are similar in clusters 1 and 2. Similarly, the $p$-value of the Chow test between clusters 1 and 2 is greater than 0.05, indicating that they likely come from the same regression model. This may be because both clusters are repeater clusters. The right panel of Figure \ref{fig:ker} displays the empirical relation between two other independent parameters: spectral acceleration ($r$) and spectral index ($\gamma$). In contrast to the ${\rm log \,}\Delta t_{rw} - {\rm log \,}\Delta t_{sc}$ relation, the $r - \gamma$ relation is less evident in cluster 2, but is very pronounced in clusters 0 and 1 ($R^2>0.8$), with similar slopes and intercepts. Moreover, the $p$-value of the Chow test between these two clusters even reaches around 0.5, indicating that some repeaters may share spectral properties with non-repeaters. Combining the two repeater clusters, cluster 1 and cluster 2, the relation ${\rm log \,}\Delta t_{sc} - {\rm log \,}\Delta t_{rw}$ yields an $R^2$ value of 0.4608. Although this is less than 0.5, it still exhibits a distinct empirical correlation compared to non-repeating bursts. The $r - \gamma$ relation for the combined clusters has an $R^2$ value of 0.3767, which is even lower than that of cluster 2. This may result from the substantial disparity between the $r - \gamma$ relations of cluster 1 and cluster 2, as well as the relatively weak correlation observed in cluster 2. The $p$-values of the Chow test between the non-repeater cluster (Cluster 0) and the repeater clusters (Clusters 1 and 2) are all below 0.05 for both empirical relations. This result strengthens the idea that FRBs can be classified into repeaters and non-repeaters.

Figure \ref{fig:her} shows the empirical relations for different clusters from HDBSCAN, and Table \ref{tab:her} lists the slope, intercept, and score of these empirical relations. The left panel displays the relation between rest-frame width (${\rm log \,}\Delta t_{sc}$) and the brightness temperature (${\rm log \,}T_{B}$), which are also two independent parameters. We can see that only cluster 0 (a non-repeater cluster) has a significant ${\rm log \,}\Delta t_{sc} - {\rm log \,}T_{B}$ relation with $R^2 > 0.6$, while the other clusters show no strong ${\rm log \,}\Delta t_{sc} - {\rm log \,}T_{B}$ relation. Even if the $p$-values of the Chow test between some clusters are greater than 0.05, it does not necessarily indicate that they originate from the same regression model. The right panel shows the empirical relation between spectral acceleration ($r$) and spectral index ($\gamma$). All clusters exhibit a significant $r - \gamma$ relation, especially clusters 0-3, where $R^2 > 0.8$. Notably, clusters 0 and 3 (a repeater cluster) have very similar slopes and intercepts for the $r - \gamma$ relation (the $p$-value of the Chow test is 0.45), suggesting that certain repeaters might have spectral properties similar to those of non-repeaters, a trend also observed in the k-means results. We also analyzed the relations using the combined data from clusters 0 and 1 (non-repeater clusters), as well as from clusters 2, 3, and 4 (repeater clusters). For the ${\rm log \,}\Delta t_{sc} - {\rm log \,}T_{B}$ relation, the $R^2$ value for the non-repeater clusters is less than 0.5, likely due to the scattered data in cluster 1. However, it still performs better than the repeater cluster, which has an $R^2$ value of 0.0444, indicating that the ${\rm log \,}\Delta t_{sc} - {\rm log \,}T_{B}$ relation is virtually non-existent for the repeater cluster. The $r - \gamma$ relation is still apparent in the non-repeater clusters, but the $R^2$ value is less than 0.5 in the repeater clusters. This may be due to the significant differences in slope and intercept of the $r - \gamma$ relation across clusters 2, 3, and 4. The $p$-values of the Chow test between non-repeater clusters and repeater clusters are all significantly below 0.05 for both empirical relations, aligning with the results from k-means clustering.

\section{Conclusion and Future Prospect} \label{sec:con}
Machine learning is a powerful tool to classify FRBs. In this paper, we applied unsupervised learning methods, including dimensionality reduction and clustering algorithms, to differentiate between repeaters and non-repeaters in the first CHIME/FRB catalog \citep{CHIMEFRBcat1} and the CHIME/FRB Collaboration (2023) catalog \citep{CHIMEFRB2023}. We extracted 16 parameters from the FRBs to serve as input features for unsupervised learning, ensuring that the information from the FRBs was sufficiently comprehensive. Ultimately, we successfully identified several candidate repeaters among the non-repeaters. Using the UMAP+k-means method, we identified 269 non-repeaters as repeater candidates, with an estimated repeater source percentage of 61.7\%. With the UMAP+HDBSCAN method, 141 non-repeaters were identified as repeater candidates, yielding a repeater source percentage of 37.9\%. All repeater candidates are summarized in Appendix \ref{app:rcl}. Additionally, we found that FRBs in repeater clusters and non-repeater clusters exhibit different distributions across several features, suggesting that repeaters and non-repeaters may belong to distinct categories.

We used six FRB sources previously classified as non-repeaters but actually confirmed as repeaters to evaluate the predictive capability of our model. The UMAP + k-means method successfully predicted five of these sources, while the UMAP + HDBSCAN method successfully predicted four. The only exception was FRB\,20180910A, which could not be predicted. The reason for this is that many of its characteristics, such as frequency bandwidth, spectral index, and spectral running, closely resemble those of non-repeaters, making it distinctly different from typical repeaters. Additionally, the intervals between the three outbursts of this repeater are quite long, and the features of each burst show significant variation. This may indicate that the FRB sources are unrelated and originate from different galaxies within the same direction or from different sources in the same galaxy. Furthermore, within Cat1, there are some localized non-repeaters, and we identified two of them as repeater candidates using both clustering algorithms. Continued observations of the host galaxies of these two FRBs may reveal additional repeating bursts in the future.

We further analyzed the empirical relations that may exist within different clusters. For the clusters derived from k-means, we identified a significant (with $R^2>0.5$) ${\rm log \,}\Delta t_{sc} - {\rm log \,}\Delta t_{rw}$ relation exclusive to cluster 2, as well as a $r - \gamma$ relation present only in clusters 0 and 1. For the clusters obtained through HDBSCAN, we found a notable ${\rm log \,}\Delta t_{sc} - {\rm log \,}T_B$ relation that exists solely in cluster 0, along with the $r - \gamma$ relation observed across all clusters. The spectral index $\gamma$ and the spectral running $r$ are the shape parameters of the FRB spectrum, described by a continuous power-law function \citep{FRBmophology2021ApJ, Planck:2018vyg}:
\begin{equation}\label{eq:fs}
    I(\nu) = A (\nu/\nu_0)^{\gamma + r\,{\rm ln}(\nu/\nu_0)},
\end{equation}
where $I(\nu)$ is the intensity at spectral frequency $\nu$,
A is the amplitude, and $\nu_0$ is the pivotal frequency, set at 400.1953125 MHz, the lower limit of the CHIME band. The strict $r - \gamma$ relation means that only one parameter can determine the morphology of FRB. Some clusters from k-means exhibit relatively weak but still noteworthy empirical relations, such as the ${\rm log \,}\Delta t_{sc} - {\rm log \,}\Delta t_{rw}$ relation in clusters 0 and 1 and the $r - \gamma$ relation in cluster 2. Similarly, for HDBSCAN, the ${\rm log \,}\Delta t_{sc} - {\rm log \,}T_B$ relation is present but weak in cluster 1, while it is extremely weak in clusters 2, 3, and 4, suggesting that it is effectively nonexistent in the latter. These results suggest that the ${\rm log \,}\Delta t_{sc} - {\rm log \,}\Delta t_{rw}$ relation is more pronounced in repeater clusters, whereas the ${\rm log \,}\Delta t_{sc} - {\rm log \,}T_B$ relation is more evident in non-repeater clusters. While the $r - \gamma$ relation is significant in both repeaters and non-repeaters, the noticeable differences in slopes and intercepts across different repeater clusters indicate that repeaters do not form a homogeneous group, and their properties may vary significantly.

We also applied the Chow test to assess statistical differences in the slopes and intercepts of empirical relationships across different clusters. For certain prominent empirical relationships, the Chow test $p$-values indicate that some repeater and non-repeater clusters share a common regression model, such as the $r - \gamma$ relation in clusters 0 and 1 from k-means and clusters 0 and 3 from HDBSCAN. This suggests that some repeaters and non-repeaters exhibit comparable spectral characteristics, a finding consistent with observational evidence—particularly in the case of FRB\,20180910A, if it is indeed a genuine repeater. However, this also implies that if some repeaters behave similarly to non-repeaters in key spectral relationships, some FRBs currently classified as non-repeaters might actually be repeaters with undetected bursts due to observational limitations. Subsequently, when all repeater and non-repeater clusters are merged separately, the Chow test $p$-values reveal significantly distinct empirical relationships between the two groups, reinforcing the notion that FRBs can indeed be categorized into repeaters and non-repeaters.

In the future, improvements in observational data and machine learning techniques will refine FRB classification, reduce misclassifications, and uncover more repeaters. The use of advanced clustering algorithms and multi-wavelength observations will enhance the accuracy of models, while deep learning approaches may reveal new patterns. These advancements will contribute to a better understanding of the origins and physical mechanisms of FRBs, with potential implications for cosmology.

\begin{acknowledgments}
We are grateful to Hao Wei, Jing-Yi Jia, Lin-Yu Li, Jia-Lei Niu, and Yu-Xuan Li for their kind assistance and valuable discussions. D.-C.Q. is supported by the Startup Research Fund of Henan Academy of Sciences No.241841222. S.Y. acknowledges the funding from the National Natural Science Foundation of China under grant No. 12303046, and the Startup Research Fund of Henan Academy of Sciences No.241841217. J.Z. is supported by the funding from the National Natural Science Foundation of China under grant No.12403002, and the Startup Research Fund of Henan Academy of Sciences No.241841221. Z.-Q.Y. is supported by the National Natural Science Foundation of China under grant No. 12305059, the Joint Fund of Henan Province Science and Technology R\&D Program No.235200810111, and the Startup Research Fund of Henan Academy of Sciences No.241841224. 

\end{acknowledgments}

\bibliography{reffrb}{}

\begin{thebibliography}{}
\expandafter\ifx\csname natexlab\endcsname\relax\def\natexlab#1{#1}\fi
\providecommand{\url}[1]{\href{#1}{#1}}
\providecommand{\dodoi}[1]{doi:~\href{http://doi.org/#1}{\nolinkurl{#1}}}
\providecommand{\doeprint}[1]{\href{http://ascl.net/#1}{\nolinkurl{http://ascl.net/#1}}}
\providecommand{\doarXiv}[1]{\href{https://arxiv.org/abs/#1}{\nolinkurl{https://arxiv.org/abs/#1}}}

\bibitem[{{Ad{\'a}mek} \& {Armour}(2020)}]{Adamek2020ApJS}
{Ad{\'a}mek}, K., \& {Armour}, W. 2020, \apjs, 247, 56, \dodoi{10.3847/1538-4365/ab7994}

\bibitem[{{Agarwal} {et~al.}(2020){Agarwal}, {Aggarwal}, {Burke-Spolaor}, {Lorimer}, \& {Garver-Daniels}}]{Agarwal2020MNRAS}
{Agarwal}, D., {Aggarwal}, K., {Burke-Spolaor}, S., {Lorimer}, D.~R., \& {Garver-Daniels}, N. 2020, \mnras, 497, 1661, \dodoi{10.1093/mnras/staa1856}

\bibitem[{{Bhardwaj} {et~al.}(2024){Bhardwaj}, {Michilli}, {Kirichenko}, {Modilim}, {Shin}, {Kaspi}, {Andersen}, {Cassanelli}, {Brar}, {Chatterjee}, {Cook}, {Dong}, {Fonseca}, {Gaensler}, {Ibik}, {Kaczmarek}, {Lanman}, {Leung}, {Masui}, {Pandhi}, {Pearlman}, {Petroff}, {Pleunis}, {Prochaska}, {Rafiei-Ravandi}, {Sand}, {Scholz}, \& {Smith}}]{Bhardwaj2024ApJ}
{Bhardwaj}, M., {Michilli}, D., {Kirichenko}, A.~Y., {et~al.} 2024, \apjl, 971, L51, \dodoi{10.3847/2041-8213/ad64d1}

\bibitem[{{Bhatporia} {et~al.}(2023){Bhatporia}, {Walters}, {Murugan}, \& {Weltman}}]{Bhatporia2023}
{Bhatporia}, S., {Walters}, A., {Murugan}, J., \& {Weltman}, A. 2023, arXiv e-prints, arXiv:2311.03456, \dodoi{10.48550/arXiv.2311.03456}

\bibitem[{Breiman(2001)}]{ml2}
Breiman, L. 2001, Machine Learning, 45, 5, \dodoi{10.1023/A:1010933404324}

\bibitem[{Burke-Spolaor \& Bannister(2014)}]{Burke-Spolaor:2014rqa}
Burke-Spolaor, S., \& Bannister, K.~W. 2014, Astrophys. J., 792, 19, \dodoi{10.1088/0004-637X/792/1/19}

\bibitem[{Campello {et~al.}(2013)Campello, Moulavi, \& Sander}]{Campello_2013_HDBSCAN}
Campello, R. J. G.~B., Moulavi, D., \& Sander, J. 2013, in Advances in Knowledge Discovery and Data Mining, ed. J.~Pei, V.~S. Tseng, L.~Cao, H.~Motoda, \& G.~Xu (Berlin, Heidelberg: Springer Berlin Heidelberg), 160--172

\bibitem[{Campello {et~al.}(2015)Campello, Moulavi, Zimek, \& Sander}]{Campello2015-HDBSCAN}
Campello, R. J. G.~B., Moulavi, D., Zimek, A., \& Sander, J. 2015, ACM Trans. Knowl. Discov. Data, 10, \dodoi{10.1145/2733381}

\bibitem[{Chang \& Lin(2011)}]{ml3}
Chang, C.-C., \& Lin, C.-J. 2011, ACM Trans. Intell. Syst. Technol., 2, \dodoi{10.1145/1961189.1961199}

\bibitem[{Chen {et~al.}(2021)Chen, Hashimoto, Goto, Kim, Santos, On, Lu, \& Hsiao}]{Chen:2021jpq}
Chen, B.~H., Hashimoto, T., Goto, T., {et~al.} 2021, Mon. Not. Roy. Astron. Soc., 509, 1227, \dodoi{10.1093/mnras/stab2994}

\bibitem[{{Chen} {et~al.}(2022){Chen}, {Hashimoto}, {Goto}, {Kim}, {Santos}, {On}, {Lu}, \& {Hsiao}}]{Chen2022}
{Chen}, B.~H., {Hashimoto}, T., {Goto}, T., {et~al.} 2022, \mnras, 509, 1227, \dodoi{10.1093/mnras/stab2994}

\bibitem[{Chen {et~al.}(2023)Chen, Hashimoto, Goto, Raquel, Uno, Kim, Hsiao, \& Ho}]{Chen:2023pub}
Chen, B.~H., Hashimoto, T., Goto, T., {et~al.} 2023, Mon. Not. Roy. Astron. Soc., 521, 5738, \dodoi{10.1093/mnras/stad930}

\bibitem[{{CHIME/FRB Collaboration} {et~al.}(2018){CHIME/FRB Collaboration}, {Amiri}, {Bandura}, {Berger}, {Bhardwaj}, {Boyce}, {Boyle}, {Brar}, {Burhanpurkar}, {Chawla}, {Chowdhury}, {Cliche}, {Cranmer}, {Cubranic}, {Deng}, {Denman}, {Dobbs}, {Fandino}, {Fonseca}, {Gaensler}, {Giri}, {Gilbert}, {Good}, {Guliani}, {Halpern}, {Hinshaw}, {H{\"o}fer}, {Josephy}, {Kaspi}, {Landecker}, {Lang}, {Liao}, {Masui}, {Mena-Parra}, {Naidu}, {Newburgh}, {Ng}, {Patel}, {Pen}, {Pinsonneault-Marotte}, {Pleunis}, {Rafiei Ravandi}, {Ransom}, {Renard}, {Scholz}, {Sigurdson}, {Siegel}, {Smith}, {Stairs}, {Tendulkar}, {Vanderlinde}, \& {Wiebe}}]{CHIMEFRB-IM}
{CHIME/FRB Collaboration}, {Amiri}, M., {Bandura}, K., {et~al.} 2018, \apj, 863, 48, \dodoi{10.3847/1538-4357/aad188}

\bibitem[{{CHIME/FRB Collaboration} {et~al.}(2021){CHIME/FRB Collaboration}, {Amiri}, {Andersen}, {Bandura}, {Berger}, {Bhardwaj}, {Boyce}, {Boyle}, {Brar}, {Breitman}, {Cassanelli}, {Chawla}, {Chen}, {Cliche}, {Cook}, {Cubranic}, {Curtin}, {Deng}, {Dobbs}, {Dong}, {Eadie}, {Fandino}, {Fonseca}, {Gaensler}, {Giri}, {Good}, {Halpern}, {Hill}, {Hinshaw}, {Josephy}, {Kaczmarek}, {Kader}, {Kania}, {Kaspi}, {Landecker}, {Lang}, {Leung}, {Li}, {Lin}, {Masui}, {McKinven}, {Mena-Parra}, {Merryfield}, {Meyers}, {Michilli}, {Milutinovic}, {Mirhosseini}, {M{\"u}nchmeyer}, {Naidu}, {Newburgh}, {Ng}, {Patel}, {Pen}, {Petroff}, {Pinsonneault-Marotte}, {Pleunis}, {Rafiei-Ravandi}, {Rahman}, {Ransom}, {Renard}, {Sanghavi}, {Scholz}, {Shaw}, {Shin}, {Siegel}, {Sikora}, {Singh}, {Smith}, {Stairs}, {Tan}, {Tendulkar}, {Vanderlinde}, {Wang}, {Wulf}, \& {Zwaniga}}]{CHIMEFRBcat1}
{CHIME/FRB Collaboration}, {Amiri}, M., {Andersen}, B.~C., {et~al.} 2021, \apjs, 257, 59, \dodoi{10.3847/1538-4365/ac33ab}

\bibitem[{{CHIME/FRB Collaboration} {et~al.}(2023){CHIME/FRB Collaboration}, {Andersen}, {Bandura}, {Bhardwaj}, {Boyle}, {Brar}, {Cassanelli}, {Chatterjee}, {Chawla}, {Cook}, {Curtin}, {Dobbs}, {Dong}, {Faber}, {Fandino}, {Fonseca}, {Gaensler}, {Giri}, {Herrera-Martin}, {Hill}, {Ibik}, {Josephy}, {Kaczmarek}, {Kader}, {Kaspi}, {Landecker}, {Lanman}, {Lazda}, {Leung}, {Lin}, {Masui}, {McKinven}, {Mena-Parra}, {Meyers}, {Michilli}, {Ng}, {Pandhi}, {Pearlman}, {Pen}, {Petroff}, {Pleunis}, {Rafiei-Ravandi}, {Rahman}, {Ransom}, {Renard}, {Sand}, {Sanghavi}, {Scholz}, {Shah}, {Shin}, {Siegel}, {Smith}, {Stairs}, {Su}, {Tendulkar}, {Vanderlinde}, {Wang}, {Wulf}, \& {Zwaniga}}]{CHIMEFRB2023}
{CHIME/FRB Collaboration}, {Andersen}, B.~C., {Bandura}, K., {et~al.} 2023, \apj, 947, 83, \dodoi{10.3847/1538-4357/acc6c1}

\bibitem[{Chow(1960)}]{chowtest}
Chow, G.~C. 1960, Econometrica, 28, 591.
\newblock \url{http://www.jstor.org/stable/1910133}

\bibitem[{{Cordes} \& {Lazio}(2002)}]{ne2001}
{Cordes}, J.~M., \& {Lazio}, T.~J.~W. 2002, arXiv e-prints, astro, \dodoi{10.48550/arXiv.astro-ph/0207156}

\bibitem[{Cover \& Hart(1967)}]{ml1967}
Cover, T., \& Hart, P. 1967, IEEE Trans. Inf. Theor., 13, 21–27, \dodoi{10.1109/TIT.1967.1053964}

\bibitem[{Dempster {et~al.}(2018)Dempster, Laird, \& Rubin}]{ml1}
Dempster, A.~P., Laird, N.~M., \& Rubin, D.~B. 2018, Journal of the Royal Statistical Society: Series B (Methodological), 39, 1, \dodoi{10.1111/j.2517-6161.1977.tb01600.x}

\bibitem[{Deng \& Zhang(2014)}]{Deng:2013aga}
Deng, W., \& Zhang, B. 2014, Astrophys. J. Lett., 783, L35, \dodoi{10.1088/2041-8205/783/2/L35}

\bibitem[{Ester {et~al.}(1996)Ester, Kriegel, Sander, Xu, {et~al.}}]{ester1996_dbscan}
Ester, M., Kriegel, H.-P., Sander, J., Xu, X., {et~al.} 1996in , 226--231

\bibitem[{{Fonseca} {et~al.}(2020){Fonseca}, {Andersen}, {Bhardwaj}, {Chawla}, {Good}, {Josephy}, {Kaspi}, {Masui}, {Mckinven}, {Michilli}, {Pleunis}, {Shin}, {Tendulkar}, {Bandura}, {Boyle}, {Brar}, {Cassanelli}, {Cubranic}, {Dobbs}, {Dong}, {Gaensler}, {Hinshaw}, {Landecker}, {Leung}, {Li}, {Lin}, {Mena-Parra}, {Merryfield}, {Naidu}, {Ng}, {Patel}, {Pen}, {Rafiei-Ravandi}, {Rahman}, {Ransom}, {Scholz}, {Smith}, {Stairs}, {Vanderlinde}, {Yadav}, \& {Zwaniga}}]{Fonseca:2020cdd}
{Fonseca}, E., {Andersen}, B.~C., {Bhardwaj}, M., {et~al.} 2020, \apjl, 891, L6, \dodoi{10.3847/2041-8213/ab7208}

\bibitem[{{Fotopoulou}(2024)}]{umlrevi}
{Fotopoulou}, S. 2024, Astronomy and Computing, 48, 100851, \dodoi{10.1016/j.ascom.2024.100851}

\bibitem[{Gao {et~al.}(2014)Gao, Li, \& Zhang}]{Gao:2014iva}
Gao, H., Li, Z., \& Zhang, B. 2014, Astrophys. J., 788, 189, \dodoi{10.1088/0004-637X/788/2/189}

\bibitem[{{Gordon} {et~al.}(2023){Gordon}, {Fong}, {Kilpatrick}, {Eftekhari}, {Leja}, {Prochaska}, {Nugent}, {Bhandari}, {Blanchard}, {Caleb}, {Day}, {Deller}, {Dong}, {Glowacki}, {Gourdji}, {Mannings}, {Mahoney}, {Marnoch}, {Miller}, {Paterson}, {Rastinejad}, {Ryder}, {Sadler}, {Scott}, {Sears}, {Shannon}, {Simha}, {Stappers}, \& {Tejos}}]{Gordon2023ApJ}
{Gordon}, A.~C., {Fong}, W.-f., {Kilpatrick}, C.~D., {et~al.} 2023, \apj, 954, 80, \dodoi{10.3847/1538-4357/ace5aa}

\bibitem[{Guo \& Wei(2022)}]{Guo:2022wpf}
Guo, H.-Y., \& Wei, H. 2022, JCAP, 07, 010, \dodoi{10.1088/1475-7516/2022/07/010}

\bibitem[{Guo \& Wei(2024)}]{Guo:2023hgb}
---. 2024, Phys. Lett. B, 859, 139120, \dodoi{10.1016/j.physletb.2024.139120}

\bibitem[{{Hallinan} {et~al.}(2019){Hallinan}, {Ravi}, {Weinreb}, {Kocz}, {Huang}, {Woody}, {Lamb}, {D'Addario}, {Catha}, {Law}, {Kulkarni}, {Phinney}, {Eastwood}, {Bouman}, {McLaughlin}, {Ransom}, {Siemens}, {Cordes}, {Lynch}, {Kaplan}, {Brazier}, {Bhatnagar}, {Myers}, {Walter}, \& {Gaensler}}]{DSA-2000}
{Hallinan}, G., {Ravi}, V., {Weinreb}, S., {et~al.} 2019, in Bulletin of the American Astronomical Society, Vol.~51, 255, \dodoi{10.48550/arXiv.1907.07648}

\bibitem[{Han {et~al.}(2012)Han, Kamber, \& Pei}]{HAN2012443}
Han, J., Kamber, M., \& Pei, J. 2012, in Data Mining (Third Edition), third edition edn., ed. J.~Han, M.~Kamber, \& J.~Pei, The Morgan Kaufmann Series in Data Management Systems (Boston: Morgan Kaufmann), 443--495, \dodoi{https://doi.org/10.1016/B978-0-12-381479-1.00010-1}

\bibitem[{{Jankowski} {et~al.}(2023){Jankowski}, {Bezuidenhout}, {Caleb}, {Driessen}, {Malenta}, {Morello}, {Rajwade}, {Sanidas}, {Stappers}, {Surnis}, {Barr}, {Chen}, {Kramer}, {Wu}, {Buchner}, {Serylak}, \& {Prochaska}}]{Jankowski:2023sot}
{Jankowski}, F., {Bezuidenhout}, M.~C., {Caleb}, M., {et~al.} 2023, \mnras, 524, 4275, \dodoi{10.1093/mnras/stad2041}

\bibitem[{Keane(2018)}]{Keane:2018jqo}
Keane, E.~F. 2018, Nature Astron., 2, 865, \dodoi{10.1038/s41550-018-0603-0}

\bibitem[{{Kirsten} {et~al.}(2022){Kirsten}, {Marcote}, {Nimmo}, {Hessels}, {Bhardwaj}, {Tendulkar}, {Keimpema}, {Yang}, {Snelders}, {Scholz}, {Pearlman}, {Law}, {Peters}, {Giroletti}, {Paragi}, {Bassa}, {Hewitt}, {Bach}, {Bezrukovs}, {Burgay}, {Buttaccio}, {Conway}, {Corongiu}, {Feiler}, {Forss{\'e}n}, {Gawro{\'n}ski}, {Karuppusamy}, {Kharinov}, {Lindqvist}, {Maccaferri}, {Melnikov}, {Ould-Boukattine}, {Possenti}, {Surcis}, {Wang}, {Yuan}, {Aggarwal}, {Anna-Thomas}, {Bower}, {Blaauw}, {Burke-Spolaor}, {Cassanelli}, {Clarke}, {Fonseca}, {Gaensler}, {Gopinath}, {Kaspi}, {Kassim}, {Lazio}, {Leung}, {Li}, {Lin}, {Masui}, {Mckinven}, {Michilli}, {Mikhailov}, {Ng}, {Orbidans}, {Pen}, {Petroff}, {Rahman}, {Ransom}, {Shin}, {Smith}, {Stairs}, \& {Vlemmings}}]{Kirsten:2021llv}
{Kirsten}, F., {Marcote}, B., {Nimmo}, K., {et~al.} 2022, \nat, 602, 585, \dodoi{10.1038/s41586-021-04354-w}

\bibitem[{{Kocz} {et~al.}(2019){Kocz}, {Ravi}, {Catha}, {D'Addario}, {Hallinan}, {Hobbs}, {Kulkarni}, {Shi}, {Vedantham}, {Weinreb}, \& {Woody}}]{DSA-10}
{Kocz}, J., {Ravi}, V., {Catha}, M., {et~al.} 2019, \mnras, 489, 919, \dodoi{10.1093/mnras/stz2219}

\bibitem[{Kumar \& Zhang(2014)}]{grb2}
Kumar, P., \& Zhang, B. 2014, Phys. Rept., 561, 1, \dodoi{10.1016/j.physrep.2014.09.008}

\bibitem[{{Kumar} {et~al.}(2019){Kumar}, {Shannon}, {Os{\l}owski}, {Qiu}, {Bhandari}, {Farah}, {Flynn}, {Kerr}, {Lorimer}, {Macquart}, {Ng}, {Phillips}, {Price}, \& {Spiewak}}]{Kumar2019ApJ}
{Kumar}, P., {Shannon}, R.~M., {Os{\l}owski}, S., {et~al.} 2019, \apjl, 887, L30, \dodoi{10.3847/2041-8213/ab5b08}

\bibitem[{{Law} {et~al.}(2020){Law}, {Butler}, {Prochaska}, {Zackay}, {Burke-Spolaor}, {Mannings}, {Tejos}, {Josephy}, {Andersen}, {Chawla}, {Heintz}, {Aggarwal}, {Bower}, {Demorest}, {Kilpatrick}, {Lazio}, {Linford}, {Mckinven}, {Tendulkar}, \& {Simha}}]{Law2020ApJ}
{Law}, C.~J., {Butler}, B.~J., {Prochaska}, J.~X., {et~al.} 2020, \apj, 899, 161, \dodoi{10.3847/1538-4357/aba4ac}

\bibitem[{{Law} {et~al.}(2024){Law}, {Sharma}, {Ravi}, {Chen}, {Catha}, {Connor}, {Faber}, {Hallinan}, {Harnach}, {Hellbourg}, {Hobbs}, {Hodge}, {Hodges}, {Lamb}, {Rasmussen}, {Sherman}, {Shi}, {Simard}, {Squillace}, {Weinreb}, {Woody}, \& {Yurk}}]{DSA-110_cat1}
{Law}, C.~J., {Sharma}, K., {Ravi}, V., {et~al.} 2024, \apj, 967, 29, \dodoi{10.3847/1538-4357/ad3736}

\bibitem[{{Li} \& {Pan}(2016)}]{FAST-IM}
{Li}, D., \& {Pan}, Z. 2016, Radio Science, 51, 1060, \dodoi{10.1002/2015RS005877}

\bibitem[{{Li} {et~al.}(2024){Li}, {Jia}, {Qiang}, \& {Wei}}]{Li2024arXiv240812983}
{Li}, L.-Y., {Jia}, J.-Y., {Qiang}, D.-C., \& {Wei}, H. 2024, arXiv e-prints, arXiv:2408.12983, \dodoi{10.48550/arXiv.2408.12983}

\bibitem[{Li {et~al.}(2021)Li, Dong, Zhang, \& Li}]{Li:2021yds}
Li, X.~J., Dong, X.~F., Zhang, Z.~B., \& Li, D. 2021, Astrophys. J., 923, 230, \dodoi{10.3847/1538-4357/ac3085}

\bibitem[{Li {et~al.}(2019)Li, Gao, Wei, Yang, Zhang, \& Zhu}]{Li:2019klc}
Li, Z., Gao, H., Wei, J.-J., {et~al.} 2019, Astrophys. J., 876, 146, \dodoi{10.3847/1538-4357/ab18fe}

\bibitem[{Lloyd(1982)}]{lloyd1982_k-means}
Lloyd, S. 1982, IEEE transactions on information theory, 28, 129

\bibitem[{Lorimer(2018)}]{Lorimer:2018rwi}
Lorimer, D.~R. 2018, Nature Astron., 2, 860, \dodoi{10.1038/s41550-018-0607-9}

\bibitem[{Lorimer {et~al.}(2007)Lorimer, Bailes, McLaughlin, Narkevic, \& Crawford}]{Lorimer:2007qn}
Lorimer, D.~R., Bailes, M., McLaughlin, M.~A., Narkevic, D.~J., \& Crawford, F. 2007, Science, 318, 777, \dodoi{10.1126/science.1147532}

\bibitem[{Luo {et~al.}(2022)Luo, Zhu-Ge, \& Zhang}]{Luo:2022smj}
Luo, J.-W., Zhu-Ge, J.-M., \& Zhang, B. 2022, Mon. Not. Roy. Astron. Soc., 518, 1629, \dodoi{10.1093/mnras/stac3206}

\bibitem[{MacQueen {et~al.}(1967)}]{macqueen1967_k-means}
MacQueen, J., {et~al.} 1967in , Oakland, CA, USA, 281--297

\bibitem[{McGregor \& Lorimer(2024)}]{McGregor:2023mzr}
McGregor, K., \& Lorimer, D.~R. 2024, Astrophys. J., 961, 10, \dodoi{10.3847/1538-4357/ad1184}

\bibitem[{McInnes {et~al.}(2017)McInnes, Healy, \& Astels}]{McInnes2017_HDBSCAN}
McInnes, L., Healy, J., \& Astels, S. 2017, Journal of Open Source Software, 2, 205, \dodoi{10.21105/joss.00205}

\bibitem[{McInnes {et~al.}(2018)McInnes, Healy, \& Melville}]{umap}
McInnes, L., Healy, J., \& Melville, J. 2018.
\newblock \doarXiv{1802.03426}

\bibitem[{{Niu} {et~al.}(2022){Niu}, {Aggarwal}, {Li}, {Zhang}, {Chatterjee}, {Tsai}, {Yu}, {Law}, {Burke-Spolaor}, {Cordes}, {Zhang}, {Ocker}, {Yao}, {Wang}, {Feng}, {Niino}, {Bochenek}, {Cruces}, {Connor}, {Jiang}, {Dai}, {Luo}, {Li}, {Miao}, {Niu}, {Anna-Thomas}, {Sydnor}, {Stern}, {Wang}, {Yuan}, {Yue}, {Zhou}, {Yan}, {Zhu}, \& {Zhang}}]{Niu:2021bnl}
{Niu}, C.~H., {Aggarwal}, K., {Li}, D., {et~al.} 2022, \nat, 606, 873, \dodoi{10.1038/s41586-022-04755-5}

\bibitem[{Petroff {et~al.}(2022)Petroff, Hessels, \& Lorimer}]{Petroff:2021wug}
Petroff, E., Hessels, J. W.~T., \& Lorimer, D.~R. 2022, Astron. Astrophys. Rev., 30, 2, \dodoi{10.1007/s00159-022-00139-w}

\bibitem[{{Petroff} {et~al.}(2015){Petroff}, {Keane}, {Barr}, {Reynolds}, {Sarkissian}, {Edwards}, {Stevens}, {Brem}, {Jameson}, {Burke-Spolaor}, {Johnston}, {Bhat}, {Kudale}, \& {Bhandari}}]{Petroff:2015bua}
{Petroff}, E., {Keane}, E.~F., {Barr}, E.~D., {et~al.} 2015, \mnras, 451, 3933, \dodoi{10.1093/mnras/stv1242}

\bibitem[{Petroff {et~al.}(2016)Petroff, Barr, Jameson, Keane, Bailes, Kramer, Morello, Tabbara, \& van Straten}]{Petroff:2016tcr}
Petroff, E., Barr, E.~D., Jameson, A., {et~al.} 2016, Publ. Astron. Soc. Austral., 33, e045, \dodoi{10.1017/pasa.2016.35}

\bibitem[{{Planck Collaboration} {et~al.}(2020){Planck Collaboration}, {Aghanim}, {Akrami}, {Ashdown}, {Aumont}, {Baccigalupi}, {Ballardini}, {Banday}, {Barreiro}, {Bartolo}, {Basak}, {Battye}, {Benabed}, {Bernard}, {Bersanelli}, {Bielewicz}, {Bock}, {Bond}, {Borrill}, {Bouchet}, {Boulanger}, {Bucher}, {Burigana}, {Butler}, {Calabrese}, {Cardoso}, {Carron}, {Challinor}, {Chiang}, {Chluba}, {Colombo}, {Combet}, {Contreras}, {Crill}, {Cuttaia}, {de Bernardis}, {de Zotti}, {Delabrouille}, {Delouis}, {Di Valentino}, {Diego}, {Dor{\'e}}, {Douspis}, {Ducout}, {Dupac}, {Dusini}, {Efstathiou}, {Elsner}, {En{\ss}lin}, {Eriksen}, {Fantaye}, {Farhang}, {Fergusson}, {Fernandez-Cobos}, {Finelli}, {Forastieri}, {Frailis}, {Fraisse}, {Franceschi}, {Frolov}, {Galeotta}, {Galli}, {Ganga}, {G{\'e}nova-Santos}, {Gerbino}, {Ghosh}, {Gonz{\'a}lez-Nuevo}, {G{\'o}rski}, {Gratton}, {Gruppuso}, {Gudmundsson}, {Hamann}, {Handley}, {Hansen}, {Herranz}, {Hildebrandt}, {Hivon}, {Huang}, {Jaffe}, {Jones}, {Karakci}, {Keih{\"a}nen},
  {Keskitalo}, {Kiiveri}, {Kim}, {Kisner}, {Knox}, {Krachmalnicoff}, {Kunz}, {Kurki-Suonio}, {Lagache}, {Lamarre}, {Lasenby}, {Lattanzi}, {Lawrence}, {Le Jeune}, {Lemos}, {Lesgourgues}, {Levrier}, {Lewis}, {Liguori}, {Lilje}, {Lilley}, {Lindholm}, {L{\'o}pez-Caniego}, {Lubin}, {Ma}, {Mac{\'\i}as-P{\'e}rez}, {Maggio}, {Maino}, {Mandolesi}, {Mangilli}, {Marcos-Caballero}, {Maris}, {Martin}, {Martinelli}, {Mart{\'\i}nez-Gonz{\'a}lez}, {Matarrese}, {Mauri}, {McEwen}, {Meinhold}, {Melchiorri}, {Mennella}, {Migliaccio}, {Millea}, {Mitra}, {Miville-Desch{\^e}nes}, {Molinari}, {Montier}, {Morgante}, {Moss}, {Natoli}, {N{\o}rgaard-Nielsen}, {Pagano}, {Paoletti}, {Partridge}, {Patanchon}, {Peiris}, {Perrotta}, {Pettorino}, {Piacentini}, {Polastri}, {Polenta}, {Puget}, {Rachen}, {Reinecke}, {Remazeilles}, {Renzi}, {Rocha}, {Rosset}, {Roudier}, {Rubi{\~n}o-Mart{\'\i}n}, {Ruiz-Granados}, {Salvati}, {Sandri}, {Savelainen}, {Scott}, {Shellard}, {Sirignano}, {Sirri}, {Spencer}, {Sunyaev}, {Suur-Uski}, {Tauber}, {Tavagnacco},
  {Tenti}, {Toffolatti}, {Tomasi}, {Trombetti}, {Valenziano}, {Valiviita}, {Van Tent}, {Vibert}, {Vielva}, {Villa}, {Vittorio}, {Wandelt}, {Wehus}, {White}, {White}, {Zacchei}, \& {Zonca}}]{Planck:2018vyg}
{Planck Collaboration}, {Aghanim}, N., {Akrami}, Y., {et~al.} 2020, \aap, 641, A6, \dodoi{10.1051/0004-6361/201833910}

\bibitem[{{Pleunis} {et~al.}(2021){Pleunis}, {Good}, {Kaspi}, {Mckinven}, {Ransom}, {Scholz}, {Bandura}, {Bhardwaj}, {Boyle}, {Brar}, {Cassanelli}, {Chawla}, {(Adam) Dong}, {Fonseca}, {Gaensler}, {Josephy}, {Kaczmarek}, {Leung}, {Lin}, {Masui}, {Mena-Parra}, {Michilli}, {Ng}, {Patel}, {Rafiei-Ravandi}, {Rahman}, {Sanghavi}, {Shin}, {Smith}, {Stairs}, \& {Tendulkar}}]{FRBmophology2021ApJ}
{Pleunis}, Z., {Good}, D.~C., {Kaspi}, V.~M., {et~al.} 2021, \apj, 923, 1, \dodoi{10.3847/1538-4357/ac33ac}

\bibitem[{Qiang {et~al.}(2020)Qiang, Deng, \& Wei}]{Qiang:2019zrs}
Qiang, D.-C., Deng, H.-K., \& Wei, H. 2020, Class. Quant. Grav., 37, 185022, \dodoi{10.1088/1361-6382/ab7f8e}

\bibitem[{Qiang {et~al.}(2022)Qiang, Li, \& Wei}]{Qiang:2021ljr}
Qiang, D.-C., Li, S.-L., \& Wei, H. 2022, JCAP, 01, 040, \dodoi{10.1088/1475-7516/2022/01/040}

\bibitem[{Qiang \& Wei(2020)}]{Qiang:2020vta}
Qiang, D.-C., \& Wei, H. 2020, JCAP, 04, 023, \dodoi{10.1088/1475-7516/2020/04/023}

\bibitem[{Qiang \& Wei(2021)}]{Qiang:2021bwb}
---. 2021, Phys. Rev. D, 103, 083536, \dodoi{10.1103/PhysRevD.103.083536}

\bibitem[{Raquel {et~al.}(2023)Raquel, Hashimoto, Goto, Chen, Uno, Hsiao, Kim, \& Ho}]{Raquel:2023tgi}
Raquel, B. J.~R., Hashimoto, T., Goto, T., {et~al.} 2023, Mon. Not. Roy. Astron. Soc., 524, 1668, \dodoi{10.1093/mnras/stad1942}

\bibitem[{Ravi(2019)}]{Ravi:2019iop}
Ravi, V. 2019, Nature Astron., 3, 928, \dodoi{10.1038/s41550-019-0831-y}

\bibitem[{Ravi \& Lasky(2014)}]{Ravi:2014gxa}
Ravi, V., \& Lasky, P.~D. 2014, Mon. Not. Roy. Astron. Soc., 441, 2433, \dodoi{10.1093/mnras/stu720}

\bibitem[{Rousseeuw(1987)}]{silhouettes}
Rousseeuw, P.~J. 1987, Journal of Computational and Applied Mathematics, 20, 53, \dodoi{https://doi.org/10.1016/0377-0427(87)90125-7}

\bibitem[{Rumelhart {et~al.}(1986)Rumelhart, Hinton, \& Williams}]{ml1986}
Rumelhart, D.~E., Hinton, G.~E., \& Williams, R.~J. 1986, Nature, 323, 533, \dodoi{10.1038/323533a0}

\bibitem[{{Shannon} {et~al.}(2024){Shannon}, {Bannister}, {Bera}, {Bhandari}, {Day}, {Deller}, {Dial}, {Dobie}, {Ekers}, {Fong}, {Glowacki}, {Gordon}, {Gourdji}, {Jaini}, {James}, {Kumar}, {Mahony}, {Marnoch}, {Muller}, {Prochaska}, {Qiu}, {Ryder}, {Sadler}, {Scott}, {Tejos}, {Uttarkar}, \& {Wang}}]{ASKAP-IM}
{Shannon}, R.~M., {Bannister}, K.~W., {Bera}, A., {et~al.} 2024, arXiv e-prints, arXiv:2408.02083, \dodoi{10.48550/arXiv.2408.02083}

\bibitem[{{Sharma} {et~al.}(2024){Sharma}, {Ravi}, {Connor}, {Law}, {Ocker}, {Sherman}, {Kosogorov}, {Faber}, {Hallinan}, {Harnach}, {Hellbourg}, {Hobbs}, {Hodge}, {Hodges}, {Lamb}, {Rasmussen}, {Somalwar}, {Weinreb}, {Woody}, {Leja}, {Anand}, {Das}, {Qin}, {Rose}, {Dong}, {Miller}, \& {Yao}}]{DSA-110_cat2}
{Sharma}, K., {Ravi}, V., {Connor}, L., {et~al.} 2024, \nat, 635, 61, \dodoi{10.1038/s41586-024-08074-9}

\bibitem[{{Spitler} {et~al.}(2014){Spitler}, {Cordes}, {Hessels}, {Lorimer}, {McLaughlin}, {Chatterjee}, {Crawford}, {Deneva}, {Kaspi}, {Wharton}, {Allen}, {Bogdanov}, {Brazier}, {Camilo}, {Freire}, {Jenet}, {Karako-Argaman}, {Knispel}, {Lazarus}, {Lee}, {van Leeuwen}, {Lynch}, {Ransom}, {Scholz}, {Siemens}, {Stairs}, {Stovall}, {Swiggum}, {Venkataraman}, {Zhu}, {Aulbert}, \& {Fehrmann}}]{Spitler:2014fla}
{Spitler}, L.~G., {Cordes}, J.~M., {Hessels}, J.~W.~T., {et~al.} 2014, \apj, 790, 101, \dodoi{10.1088/0004-637X/790/2/101}

\bibitem[{{Spitler} {et~al.}(2016){Spitler}, {Scholz}, {Hessels}, {Bogdanov}, {Brazier}, {Camilo}, {Chatterjee}, {Cordes}, {Crawford}, {Deneva}, {Ferdman}, {Freire}, {Kaspi}, {Lazarus}, {Lynch}, {Madsen}, {McLaughlin}, {Patel}, {Ransom}, {Seymour}, {Stairs}, {Stappers}, {van Leeuwen}, \& {Zhu}}]{Spitler:2016dmz}
{Spitler}, L.~G., {Scholz}, P., {Hessels}, J.~W.~T., {et~al.} 2016, \nat, 531, 202, \dodoi{10.1038/nature17168}

\bibitem[{Sun {et~al.}(2024)Sun, Zhang, Li, Hou, Zhang, Zhang, \& Zhang}]{Sun:2024huw}
Sun, W.-P., Zhang, J.-G., Li, Y., {et~al.} 2024.
\newblock \doarXiv{2409.11173}

\bibitem[{{Thornton} {et~al.}(2013){Thornton}, {Stappers}, {Bailes}, {Barsdell}, {Bates}, {Bhat}, {Burgay}, {Burke-Spolaor}, {Champion}, {Coster}, {D'Amico}, {Jameson}, {Johnston}, {Keith}, {Kramer}, {Levin}, {Milia}, {Ng}, {Possenti}, \& {van Straten}}]{Thornton:2013iua}
{Thornton}, D., {Stappers}, B., {Bailes}, M., {et~al.} 2013, Science, 341, 53, \dodoi{10.1126/science.1236789}

\bibitem[{Vapnik(1999)}]{ml4}
Vapnik, V. 1999, IEEE Transactions on Neural Networks, 10, 988, \dodoi{10.1109/72.788640}

\bibitem[{{Wagstaff} {et~al.}(2016){Wagstaff}, {Tang}, {Thompson}, {Khudikyan}, {Wyngaard}, {Deller}, {Palaniswamy}, {Tingay}, \& {Wayth}}]{Wagstaff2016PASP}
{Wagstaff}, K.~L., {Tang}, B., {Thompson}, D.~R., {et~al.} 2016, \pasp, 128, 084503, \dodoi{10.1088/1538-3873/128/966/084503}

\bibitem[{Wang \& Wei(2023)}]{Wang:2022ami}
Wang, B., \& Wei, J.-J. 2023, Astrophys. J., 944, 50, \dodoi{10.3847/1538-4357/acb2c8}

\bibitem[{Wang {et~al.}(2020{\natexlab{a}})Wang, Wang, Yang, Yu, Zuo, \& Dai}]{Wang:2020aut}
Wang, F.~Y., Wang, Y.~Y., Yang, Y.-P., {et~al.} 2020{\natexlab{a}}, Astrophys. J., 891, 72, \dodoi{10.3847/1538-4357/ab74d0}

\bibitem[{Wang \& Zhang(2019)}]{Wang:2019sio}
Wang, F.~Y., \& Zhang, G.~Q. 2019, Astrophys. J., 882, 108, \dodoi{10.3847/1538-4357/ab35dc}

\bibitem[{Wang {et~al.}(2020{\natexlab{b}})Wang, Xu, \& Chen}]{Wang:2020rkl}
Wang, W.-Y., Xu, R., \& Chen, X. 2020{\natexlab{b}}, Astrophys. J., 899, 109, \dodoi{10.3847/1538-4357/aba268}

\bibitem[{Wei {et~al.}(2019)Wei, Li, Gao, \& Wu}]{Wei:2019uhh}
Wei, J.-J., Li, Z., Gao, H., \& Wu, X.-F. 2019, JCAP, 09, 039, \dodoi{10.1088/1475-7516/2019/09/039}

\bibitem[{{Wu} {et~al.}(2019){Wu}, {Cao}, {Lv}, {Fan}, {Tan}, \& {Yang}}]{Wu2019ApJ}
{Wu}, D., {Cao}, H., {Lv}, N., {et~al.} 2019, \apjl, 887, L10, \dodoi{10.3847/2041-8213/ab595e}

\bibitem[{Xiao \& Dai(2022)}]{Xiao:2021viy}
Xiao, D., \& Dai, Z.-G. 2022, Astron. Astrophys., 657, L7, \dodoi{10.1051/0004-6361/202142268}

\bibitem[{Xiao {et~al.}(2021)Xiao, Wang, \& Dai}]{Xiao:2021omr}
Xiao, D., Wang, F., \& Dai, Z. 2021, Sci. China Phys. Mech. Astron., 64, 249501, \dodoi{10.1007/s11433-020-1661-7}

\bibitem[{Xiao {et~al.}(2022)Xiao, Wang, \& Dai}]{Xiao:2022bbj}
---. 2022, Fast Radio Bursts, ed. C.~Bambi \& A.~Santangelo (Singapore: Springer Nature Singapore), 1--38, \dodoi{10.1007/978-981-16-4544-0_128-1}

\bibitem[{{Xu} {et~al.}(2022){Xu}, {Niu}, {Chen}, {Lee}, {Zhu}, {Dong}, {Zhang}, {Jiang}, {Wang}, {Xu}, {Zhang}, {Fu}, {Filippenko}, {Peng}, {Zhou}, {Zhang}, {Wang}, {Feng}, {Li}, {Brink}, {Li}, {Lu}, {Yang}, {Caballero}, {Cai}, {Chen}, {Dai}, {Djorgovski}, {Esamdin}, {Gan}, {Guhathakurta}, {Han}, {Hao}, {Huang}, {Jiang}, {Li}, {Li}, {Li}, {Li}, {Li}, {Liu}, {Luo}, {Men}, {Niu}, {Peng}, {Qian}, {Song}, {Stern}, {Stockton}, {Sun}, {Wang}, {Wang}, {Wang}, {Wang}, {Wu}, {Xiao}, {Xiong}, {Xu}, {Xu}, {Yang}, {Yang}, {Yao}, {Yi}, {Yue}, {Yu}, {Yu}, {Yuan}, {Zhang}, {Zhang}, {Zhang}, {Zhao}, {Zheng}, {Zhu}, \& {Zou}}]{Xu2022Natur}
{Xu}, H., {Niu}, J.~R., {Chen}, P., {et~al.} 2022, \nat, 609, 685, \dodoi{10.1038/s41586-022-05071-8}

\bibitem[{{Xu} {et~al.}(2023){Xu}, {Feng}, {Li}, {Wang}, {Zhang}, {Xie}, {Chen}, {Wang}, {Kang}, {Hu}, {Zheng}, {Tsai}, {Chen}, \& {Zhou}}]{Xu:2023did}
{Xu}, J., {Feng}, Y., {Li}, D., {et~al.} 2023, Universe, 9, 330, \dodoi{10.3390/universe9070330}

\bibitem[{Yamasaki {et~al.}(2023)Yamasaki, Goto, Ling, \& Hashimoto}]{Yamasaki:2023dlb}
Yamasaki, S., Goto, T., Ling, C.-T., \& Hashimoto, T. 2023, Mon. Not. Roy. Astron. Soc., 527, 11158, \dodoi{10.1093/mnras/stad3844}

\bibitem[{Yang {et~al.}(2023)Yang, Zhang, Wang, \& Wu}]{Yang:2023dcf}
Yang, X., Zhang, S.~B., Wang, J.~S., \& Wu, X.~F. 2023, Mon. Not. Roy. Astron. Soc., 522, 4342, \dodoi{10.1093/mnras/stad1304}

\bibitem[{{Yang} {et~al.}(2021){Yang}, {Zhang}, {Wang}, {Hobbs}, {Sun}, {Manchester}, {Geng}, {Russell}, {Luo}, {Tang}, {Wang}, {Wei}, {Staveley-Smith}, {Dai}, {Li}, {Yang}, \& {Wu}}]{Yang2021MNRAS}
{Yang}, X., {Zhang}, S.~B., {Wang}, J.~S., {et~al.} 2021, \mnras, 507, 3238, \dodoi{10.1093/mnras/stab2275}

\bibitem[{Yang {et~al.}(2017)Yang, Luo, Li, \& Zhang}]{Yang:2017bls}
Yang, Y.-P., Luo, R., Li, Z., \& Zhang, B. 2017, Astrophys. J. Lett., 839, L25, \dodoi{10.3847/2041-8213/aa6c2e}

\bibitem[{Yang \& Zhang(2016)}]{Yang:2016zbm}
Yang, Y.-P., \& Zhang, B. 2016, Astrophys. J. Lett., 830, L31, \dodoi{10.3847/2041-8205/830/2/L31}

\bibitem[{Zhang(2014)}]{Zhang:2013lta}
Zhang, B. 2014, Astrophys. J. Lett., 780, L21, \dodoi{10.1088/2041-8205/780/2/L21}

\bibitem[{Zhang {et~al.}(2007)Zhang, Zhang, Liang, Gehrels, Burrows, \& Meszaros}]{grb1}
Zhang, B., Zhang, B.-B., Liang, E.-W., {et~al.} 2007, Astrophys. J. Lett., 655, L25, \dodoi{10.1086/511781}

\bibitem[{Zhang \& Li(2018)}]{Zhang:2018gjb}
Zhang, M.-J., \& Li, H. 2018, Eur. Phys. J. C, 78, 460, \dodoi{10.1140/epjc/s10052-018-5953-3}

\bibitem[{Zhang \& Zhang(2022)}]{Zhang:2021kdu}
Zhang, R.~C., \& Zhang, B. 2022, Astrophys. J. Lett., 924, L14, \dodoi{10.3847/2041-8213/ac46ad}

\bibitem[{Zhang {et~al.}(2021)Zhang, Zhang, Li, \& Lorimer}]{Zhang:2020ass}
Zhang, R.~C., Zhang, B., Li, Y., \& Lorimer, D.~R. 2021, Mon. Not. Roy. Astron. Soc., 501, 157, \dodoi{10.1093/mnras/staa3537}

\bibitem[{{Zhang} {et~al.}(2018){Zhang}, {Gajjar}, {Foster}, {Siemion}, {Cordes}, {Law}, \& {Wang}}]{Zhang2018ApJ}
{Zhang}, Y.~G., {Gajjar}, V., {Foster}, G., {et~al.} 2018, \apj, 866, 149, \dodoi{10.3847/1538-4357/aadf31}

\bibitem[{Zhou {et~al.}(2014)Zhou, Li, Wang, Fan, \& Wei}]{Zhou:2014yta}
Zhou, B., Li, X., Wang, T., Fan, Y.-Z., \& Wei, D.-M. 2014, Phys. Rev. D, 89, 107303, \dodoi{10.1103/PhysRevD.89.107303}

\bibitem[{Zhu-Ge {et~al.}(2022)Zhu-Ge, Luo, \& Zhang}]{Zhu-Ge:2022nkz}
Zhu-Ge, J.-M., Luo, J.-W., \& Zhang, B. 2022, Mon. Not. Roy. Astron. Soc., 519, 1823, \dodoi{10.1093/mnras/stac3599}

\bibitem[{{Zhu-Ge} {et~al.}(2023){Zhu-Ge}, {Luo}, \& {Zhang}}]{zhuge2023}
{Zhu-Ge}, J.-M., {Luo}, J.-W., \& {Zhang}, B. 2023, \mnras, 519, 1823, \dodoi{10.1093/mnras/stac3599}

\end{thebibliography}
\bibliographystyle{aasjournal}

\appendix
\section{Condensed Tree of HDBSCAN}\label{app:hct}
\begin{figure}
    \centering
    \includegraphics[width=0.7\linewidth]{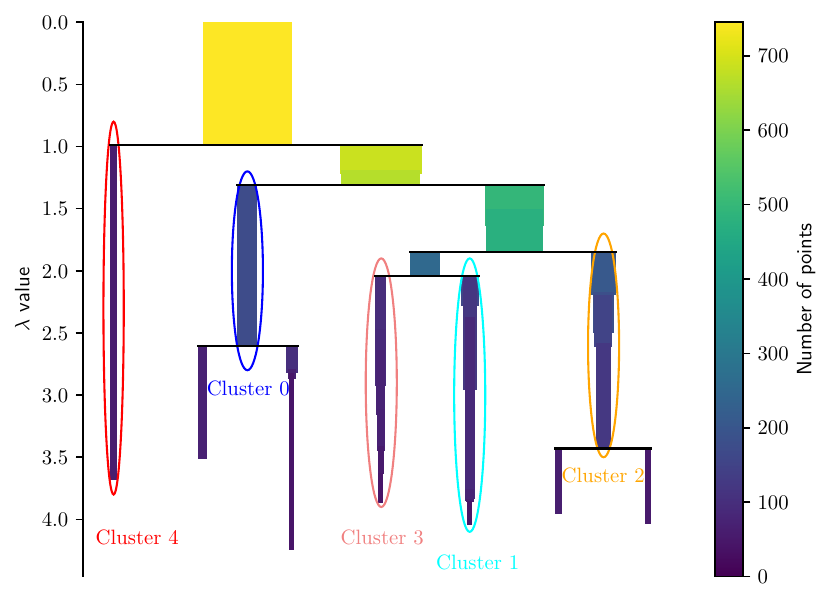}
    \caption{\label{fig:hct} The condensed tree of HDBSCAN. The leaf nodes within ellipses of different colors represent the final clusters obtained by HDBSCAN, other leaf nodes are noise.}
\end{figure}

We show the condensed tree of HDBSCAN in Figure \ref{fig:hct} to highlight the persistence and stability of the clustering results. $\lambda$ is used to consider the persistence of clusters, which can be calculated by $\lambda=1/ d$, where $d$ is the distance between a point and cluster core. For a given cluster, \(\lambda_{\mathrm{birth}}\) marks when the cluster first formed, and \(\lambda_{\mathrm{death}}\) (if applicable) marks when it split into smaller clusters. For each point \(p\) in the cluster, \(\lambda_p\) represents the value at which the point left the cluster, occurring between \(\lambda_{\mathrm{birth}}\) and \(\lambda_{\mathrm{death}}\), either during the cluster's lifetime or at the split. Then we can compute the stability for each cluster as $S = \sum_{p\in {\rm cluster}(\lambda_p - \lambda_{\rm birth})}$. Start by declaring all leaf nodes as selected clusters. Move up the tree: if the sum of child cluster stabilities exceeds the parent cluster's stability, update the parent’s stability to this sum. Otherwise, select the parent cluster and deselect its descendants. At the root node, the selected clusters form the flat clustering, which is then returned. Ultimately, we selected the five clusters enclosed within the ellipses in the Figure \ref{fig:hct}. See \url{https://hdbscan.readthedocs.io/en/latest/how_hdbscan_works.html} for more details.

\section{Repeater candidates}\label{app:rcl}
Here, we present the repeater candidates identified using unsupervised machine learning methods. The `Sub num' column lists the sub-burst numbers of FRBs from the Cat1 and Cat2023. The `Note' column specifies whether each candidate was identified exclusively by k-means (`k') or by both k-means and HDBSCAN (`both').

\centering
\startlongtable

\begin{deluxetable*}{ccccccccccccccccccccc}
\tablecaption{The list of repeater candidates \label{tab:candi_rep}}

\tablewidth{700pt}
\tabletypesize{\tiny}
\setlength{\tabcolsep}{1.5pt}
\tablehead{
\colhead{FRB} & \colhead{Sub} &\colhead{$\rm RA(J2000)$}&\colhead{$\rm DEC(J2000)$}&\colhead{$\Delta t_{bc}$}&\colhead{$\Delta t_{fitb}$}&\colhead{$S_{\nu}$}&\colhead{$F_{\nu}$}&\colhead{$\Delta t_{sc}$}&\colhead{$\gamma$}&\colhead{$r$}&\colhead{$\nu_{max}$}&\colhead{$\nu_{min}$}&\colhead{$\nu_{peak}$}&\colhead{$z$}&\colhead{${\rm log}\,E$}&\colhead{$\Delta \nu$}&\colhead{$\Delta t_{rw}$}&\colhead{${\rm log\,}L$}&\colhead{${\rm log\,}T_{B}$}&\colhead{Note}\\[-10pt]
\colhead{name}&\colhead{num}&\colhead{($^\circ$)}&\colhead{($^\circ$)}&\colhead{(ms)}&\colhead{(ms)}&\colhead{(Jy)}&\colhead{(Jy $\cdot$ ms)}&\colhead{(ms)}&\colhead{}&\colhead{}&\colhead{(MHz)}&\colhead{(MHz)}&\colhead{(MHz)}&\colhead{}&\colhead{(erg)}&\colhead{(MHz)}&\colhead{(ms)}&\colhead{(erg/s)}&\colhead{(K)}&\colhead{}\\[-14pt]
}
\startdata
FRB20180725A & 0 & 93.4 & 67.1 & 2.95 & 0.30 & 1.70 & 4.10 & 1.10 & 38.20 & -45.80 & 760.1 & 485.3 & 607.4 & 0.641 & 40.45 & 450.9 & 0.18 & 43.28 & 35.74 & k \\
FRB20180729A & 0 & 199.4 & 55.6 & 0.98 & 0.10 & 11.70 & 17.00 & 0.16 & 16.46 & -30.21 & 692.7 & 400.2 & 525.6 & 0.002 & 36.03 & 293.2 & 0.10 & 38.87 & 32.69 & k \\
FRB20180729B & 0 & 89.9 & 56.5 & 1.97 & 0.31 & 0.92 & 1.20 & 0.66 & 14.50 & -14.60 & 800.2 & 441.8 & 657.5 & 0.158 & 38.70 & 414.9 & 0.27 & 41.64 & 34.50 & k \\
FRB20180801A & 0 & 322.5 & 72.7 & 9.83 & 0.58 & 1.11 & 7.90 & 5.54 & 60.00 & -75.50 & 709.3 & 500.2 & 595.6 & 0.553 & 40.60 & 324.8 & 0.37 & 42.94 & 34.40 & both \\
FRB20180810B & 0 & 180.4 & 83.1 & 0.98 & 0.31 & 5.20 & 7.90 & 0.20 & -0.76 & -4.06 & 778.6 & 400.2 & 400.2 & 0.031 & 37.86 & 390.1 & 0.30 & 40.70 & 34.86 & k \\
FRB20180812A & 0 & 19.3 & 80.8 & 7.86 & 1.30 & 0.93 & 5.40 & 2.81 & -2.90 & -8.30 & 586.4 & 400.2 & 400.2 & 0.727 & 40.50 & 321.6 & 0.75 & 42.97 & 35.10 & both \\
FRB20180904A & 0 & 286.6 & 81.2 & 1.97 & 0.53 & 3.80 & 6.00 & 0.55 & 12.12 & -23.18 & 712.3 & 400.2 & 519.7 & 0.257 & 39.73 & 392.2 & 0.42 & 42.63 & 35.76 & k \\
FRB20180907A & 0 & 320.9 & 29.5 & 5.90 & 0.87 & 0.87 & 2.80 & 0.26 & -2.30 & -5.10 & 648.6 & 400.2 & 400.2 & 0.797 & 40.30 & 446.5 & 0.48 & 43.04 & 35.40 & k \\
\enddata
\end{deluxetable*}

\end{document}